\documentclass[12pt]{article}
\usepackage{times,amsmath,epsfig,cite}
\usepackage{multicol}
\usepackage{graphicx}
\usepackage{amsfonts}
\usepackage{amsmath}
\usepackage{framed}
\usepackage{etoolbox}
\usepackage{setspace}
\makeatletter
\preto{\@verbatim}{\topsep=0pt \partopsep=0pt }
\makeatother

\title{The Altes Family of Log-Periodic Chirplets\\and the Hyperbolic Chirplet Transform}

\author{Donnacha~Daly\thanks{Dr. Daly is corresponding author. E-mail: {\selectfont\ttfamily donnacha@ieee.org}. He conducted this work as Senior Reearcher at ETH Zurich, Switzerland. Prof. Sornette holds the Chair of Entrepreneurial Risks at ETHZ. E-mail: {\selectfont\ttfamily dsornette@ethz.ch}.} $\,$
and Didier~Sornette}

\begin{document}
\maketitle

\begin{abstract}
This work revisits a class of biomimetically inspired log-periodic waveforms first introduced by R.A. Altes in the 1970s for generalized target description. It was later observed that there is a close connection between such sonar techniques and wavelet decomposition for multiresolution analysis. Motivated by this, we formalize the original Altes waveforms as a family of hyperbolic chirplets suitable for the detection of accelerating time-series oscillations. The formalism results in a flexible set of wavelets with desirable properties of admissibility, regularity, vanishing moments, and time-frequency localization. These ``Altes wavelets'' also facilitate efficient implementation of the scale invariant hyperbolic chirplet transform (HCT).
A synthetic application is presented in this report for illustrative purposes. 
\end{abstract}

{\bf keywords}: wavelet transform, chirp, complex systems, discrete scale invariance,  critical failure, phase transition,  Doppler radar

\pagebreak
\section{Introduction}
Over the course of the 1970's, Richard A. Altes developed the theory behind a new family of waveforms with optimal Doppler tolerance for sonar applications. Inspired by mammalian acoustic echo-location calls such as those of bats and dolphins, the constructed class of time-frequency concentrated pulses consists of carefully parameterized hyperbolic chirps \cite{altes1970n1, altes1973n2, altes1975n2, altes1975n4, altes1977n2}. 
Some time later, Patrick Flandrin and his colleagues at the French national center for scientific research (CNRS) made a beautiful exposition of the close mathematical parallels between generalized target description using these chirps, and wavelet decomposition. i.e., evaluation of the wavelet transform of the target impulse response  \cite{flandrin1990generalized}. 

One of the contributions of the current work is to extend these analyses by observing that such chirps are \emph{Log-Periodic} (LP), a property closely associated with the deep symmetry of discrete scale invariance in physical systems \cite{sornette1998discrete}. By making the bridge between the chirplet approach and the LP property, we extend the toolset for detecting criticality in complex systems, with the mature body of knowledge from wavelet theory. 

Before exploring this however, it is proposed to formalize the original Altes waveforms as \emph{wavelets}. Building on the work of Flandrin \cite{flandrin1990generalized}, which considered admissibility conditions, to include other desirable wavelet properties such as the degree of regularity (smoothness), the number of vanishing moments and the degree of time-frequency (\emph{TF}) localization. The practicalities of time-discretization for efficient implementation in a Hyperbolic Chirplet Transform (HCT) are also explored.
 
Developing the HCT for detection of log-periodicty is cumbersome, when using  Altes' original formulation. As a preliminary, therefore, a re-parameterization more intuitive to the signal processing practitioner is proposed. We specify the wavelets in terms of center- and cut-off frequency (or bandwidth) and chirp-rate. In addition to familiarity, these parameters have the advantage of simplifying \emph{TF}  localization requirements and the study of other wavelet properties. The result is a powerful and practical extension to the existing body of wavelet  tools for signal analysis.

We should stress that the HCT does not claim to be a general substitute to existing wavelet transforms. Thus, our purpose is not 
to perform a comparative analysis or horse-race between the HCT and the very mature set of existing wavelet transforms 
\cite{Daubechies92,Mallat98}. Our goal is rather 
to present and study this specific Chirplet because it possesses distinct properties, namely log-periodicity resulting in some optimality characteristics
summarised below. From a practical perspective, log-periodic oscillations with an acceleration towards criticality 
can serve as indicators of an incipient bifurcation.  Such signals abound in nature, often as precursors to phase 
transitions in the non-linear dynamics of complex systems \cite{sornette1998discrete}. 
For example, the authors' interest  lies in automatic detection of the well documented phenomenon 
of log-periodic price dynamics during financial bubbles and preceding market crashes \cite{sornette2009stock}. 
However, the methodology presented here is more widely applicable in such diverse domains as prediction 
of critical failures in mechanical systems \cite{Johan-sornetterupt00}, and fault detection in electrical networks. 
Examples beyond  failure diagnostics include animal species identification via call recordings, 
commercial \& military radar, and there are probably many more.
 
The next section presents a review of relevant background material to the current work. A recap of the Altes waveform together with our re-parameterization is provided in Section~\ref{waveform}. The conditions under which it can be used as a wavelet in a HCT are presented in Section~\ref{wavelet}, along with a discussion of its wavelet properties. Section~\ref{paramSelec} examines parameter selection and wavelet design criteria, while Section~\ref{empirical} is an empirical look at the performance of the Altes wavelet in detecting noisy LP-oscillations, in comparison with other \emph{TF} methods.

\section{Transforms for Log-Periodic Chirp Detection}

\subsection{Chirplet Transforms}
\label{introchirp}

The idea of using chirps as wavelets, ie. chirplets, was introduced in \cite{mann1991chirplet, mann1995chirplet}, leading to the Gaussian chirplet transform as well as the warblet transform. A hardware implementation applied to detection of bat-calls was provided in \cite{lu2008fast}. An adaptive version of the Gaussian chirplet was designed to linearize curves in the \emph{TF}-plane in \cite{mannspieconference}. This has been used for detection of bat echo-location signals and compared with the Gaussian wavelet transform, the short time Fourier transform and the Wigner Ville Distribution (WVD) \cite{yin2002fast}, noting that

\emph{``...the Gaussian chirplet decomposition
linearizes the curving chirps. Such approximation
is quite coarse. Hyperbolic chirps probably can better
model signals, such as the sound of bat.'' }

Wavelet based LP-detection methods were also compared in \cite{sejdic2008quantitative}. The Cauchy wavelet performed worst, the Mexican Hat better and the Morlet best. More comprehensive attempts to capture general non-linear \emph{TF} modulations using the  Polynomial Chirplet Transform were developed in \cite{peng2011polynomial} and applied to bat-sonar detection in \cite{yang2013multicomponent}. None of these used a hyperbolic chirp in a wavelet transform for the detection of log-periodicity. ``Hyperbolic wavelets'' do appear elsewhere e.g. \cite{le2004hyperbolic, abry2012hyperbolic} but with a different meaning. To avoid confusion, we refer exclusively to hyperbolic chirplets in the current context.

\subsection{Power Law Chirps and the Mellin Transform}
The hyperbolic chirplet was introduced as a special case of the power law chirp in \cite{flandrin2001time}. Its use for the optimal detection of chirping gravitational waves was presented in \cite{chassande1999time}. It is seen that, while the optimal detection of a linear chirp is via the WVD, for power law chirps the optimal detector uses the Bertrand distribution, which can have prohibitive computational requirements. As an alternative, the Mellin transform can be considered, which represents a signal as a projection onto a family of hyperbolic chirps, analogous to the Fourier transform projection on a basis of complex exponentials. This idea was extended to LP-detection in \cite{poularikas2010transforms}. It was shown that the Mellin transform, coupled with a Fourier transform can be used to replace a wavelet transform, and is suited to problems of scale invariance. Similarly, in \cite{gluzman2002log} the Mellin Transform has been used to  provide the LP decomposition of the most general solution of the renormalization group equation.

\subsection{The Hyperbolic Chirplet Transform}
Here, we choose to implement the wavelet transform directly using the Altes chirp as mother wavelet. We call this a Hyperbolic Chirplet Transform (HCT), and label our parameterization of the Altes waveform as the \emph{Altes chirplet}.  Some previous studies have come close to our novel approach. 

In \cite{yiou2000dataAdaptive} a strong case, and a powerful methodology, were both presented for tailoring wavelets to the signals being detected. While the example of LP-detection was examined, the study did not extend to using a dedicated hyperbolic chirplet for the purpose. In \cite{sornette1998discrete}, a direct link was made between the functional form of the Altes wavelet (our eq.~\ref{homogeneous}) and the renormalization group encountered in \cite{saleur1996discrete, gluzman2002log}. The solution to the homogeneous Altes wavelet equation was thus identified as \emph{self-similar} with discrete scale invariance. Self similar wavelets were introduced in \cite{wornell1992wavelet} for fractal modulation.  However, these were neither identified as hyperbolic, nor used  for LP-detection. Generalized target detection was shown in \cite{flandrin1990generalized} to share important features with wavelet decomposition. This qualitative work proposed the autocorrelation function of the Altes chirp as the wavelet in the decomposition, but did not go so far as demonstrating the utility of these results. It can be considered as a useful launching point for the current work.

\subsection{Motivation: Log Periodicity \& Discrete Scale Invariance}
\label{motivation}
Apart from radar/sonar applications in target detection, there are other strong motivations for an interest in hyperbolic chirplets and their ability, via the wavelet transform, to detect log-periodic oscillations in noisy time-series. To the best of our knowledge, LP-constructs first appeared in the 1960s for modeling shocks to layered systems \cite{zababakhin1966shock} and the discrete hierarchy of vortices in hydrodynamic turbulence \cite{novikov1966effects}.
In the 1970s, log-periodicity was recognized in the self-similarity of propagating waves \cite{barenblatt1971intermediate}. Around the same time, the renormalization group theory of critical phenomena introduced solutions for the statistical mechanics of critical phase with complex critical exponents, characterized by log-periodicity \cite{jona1975renormalization,nauenberg1975scaling,van1976phase}. In the 1980s, phase transitions occurring on hierarchical lattices were shown to exhibit discrete scale invariance, with its signature of complex exponents and LP-oscillation
\cite{kapitulnik1983self,doucot1986first,bessis1987mellin,fournier1989singularity}. Since then, literature on the topic has expanded rapidly.

This potted historical perspective is condensed from \cite{sornette1998discrete} which the reader is invited to pursue for further insight. A central result is that LP-signatures indicate that a system and/or its underlying physical mechanisms have 
a hierarchy of characteristic scales corresponding to discrete scale invariance. This is interesting as it provides important insights into the underlying physics, which may allow us to make forecasts of rupture such as earthquakes \cite{sornette1995complex,saleur1996renormalization}, mechanical failure \cite{anifrani1995universal} or the bursting of bubbles in financial markets \cite{sornette1996stock, sornette2001significance, sornette2009stock}. In fact, any system with built in geometrical hierarchy will lead to log-periodicity, e.g. wave propagation in fractal systems \cite{bessis1983complex}, Ising and Potts models on hierarchical structures \cite{derrida1983fractal,meurice1995evidence} and sandpile models on discrete fractal lattices \cite{kutnjak1996sandpile} to pick a few. Given our growing understanding of the ubiquity of LP-signatures in complex systems, it is useful to equip ourselves with reliable tools for their extraction and diagnosis.

\section{Re-parameterizing the Altes Waveform}\label{waveform}
Altes' early work on chirps showed that a \emph{TF}-localized pulse which is periodic in the logarithm of progressing time has optimal Doppler tolerance \cite{altes1973n2}. We find it more instructive here to introduce the topic using his reasoning from \cite{altes1975n2}. 

\subsection{Matched-Filtering Detection of Echo Components}
Consider a waveform described in the Fourier domain by $U(\omega)$, which produces a set of echoes when reflected from a target in a sonar detection setting. Depending on the wavelength, its angle of incidence, the velocity and reflectivity of the target and other factors, reflections may be in- or out-of-phase with the incident waveform. By superposition, an echo is thus hypothesized to be the weighted sum of time-integrated (when in-phase) or -differentiated (when out-of-phase) versions of the original signal at different lags. 

The goal is to design this transmit signal $U(\omega)$ such that it can be reliably recovered from these echo components $V_n(\omega)$ under constraints on receiver complexity. Since time-integrated and -differentiated versions of $U(\omega)$ are given in the frequency domain by $(j\omega)^n U(\omega),n\in\mathbb{Z}$, detection can be achieved by using a bank of filters $V_n^*(\omega)$, each matched to one of these energy normalized echo components, written as 
\begin{equation}
V_n(\omega) = \frac{(j\omega)^n U(\omega)}
{\frac{1}{2\pi}\int_{-\infty}^{\infty}\left|(j\omega)^n U(\omega) \right|^2}.
\label{component}
\end{equation}  

\subsection{Constant-$Q$ Filter-Banks}
\label{constQsection}
Altes' insight was that the complexity of the required filter bank could be constrained if these filters have a constant time-bandwidth product for all values of $n$. This is because the component $V_n^*(\omega)$ is repeatedly differentiated as $n$ increases, which, from eq.~(\ref{component}) and the Cauchy-Schwarz inequality, results in bandwidth expansion. If this is not compensated by a corresponding compression of the time-domain impulse response, then the required matched-filter complexity grows rapidly with $n$. On the other hand, for a bank of filters with constant time-bandwidth product, also known as \emph{Constant-$Q$} \cite{vetterli1992wavelets}, filter complexity is constant. Consider matched filters $V^*_n(\omega)$ which satisfy a scaling contraint defined by\footnote{Undesirably, this $k$ introduced by Altes tunes both the bandwidth and chirp-rate of the final waveform, one of the reasons we later re-parameterize.} 
$k>1$: 
\begin{eqnarray}
V^*_n(\omega) &\propto& V^*_{n-1}\left(\frac{\omega}{k}\right).
\label{constq}
\end{eqnarray}
If the root mean square filter delay spread of $V^*_n(\omega)$ is $\tau_n$ 
and its bandwidth is $B_n$, then eq.~(\ref{constq}) gives us
$B_n = kB_{n-1}$ and its inverse Fourier transform gives $\tau_n  =  \tau_{n-1}/k$. Therefore,
\begin{equation}
\tau_nB_n = \tau_{n-1}B_{n-1}  = \tau_mB_m \;\;\forall\;\; m,n \in \mathbb{Z}.
\end{equation}
This is a sufficient constant-$Q$ condition on $V_n^*(\omega)$ and  stipulates a fixed time-bandwidth product for all $n$, as well as fixing the same ratio of center-frequency to bandwidth for all filters. From eq.~(\ref{constq}) 
\begin{eqnarray}
V^*_n(\omega)&\propto &V^*_0\left(\frac{\omega}{k^n}\right)\\
\Rightarrow \omega^nU(\omega) & \stackrel{(\ref{component})}{\propto}  & U\left(\frac{\omega}{k^n}\right)\\
\Rightarrow \omega^nU(\omega) & =  & C(n)U\left(\frac{\omega}{k^n}\right),\label{homogeneous}
\end{eqnarray}
where $C(n)$ is a proportionality constant dependent on $n$ but independent of $\omega$. From relatively simple arguments, we get to eq.~(\ref{homogeneous}), an explicitly solvable homogeneous functional in $U(\omega)$.

\subsection{The Family of Altes Waveforms}
Performing analytic continuation from integer to real values of $n$, then taking the derivative with respect to $n$ of eq.~(\ref{homogeneous}) at $n=0$ and noting that $C(0)=1$, we get 
\begin{eqnarray}
\frac{\textrm{d}\,U(\omega)}{U(\omega)}& = & \frac{C'(0)-\log{\omega}}{\omega\log k} \textrm{d}\,\omega,\;\;\omega\ne0. \label{analytic}
\end{eqnarray}
Integration and some simplification leads to the pulse $U(\omega)$ designed specifically to have echo components that are easily detectable. This is the original Altes waveform, 
\begin{equation}
U(\omega) = A \omega^\nu \exp \left(  -\frac{1}{2}\frac{\log^2\omega}{\log{k}} \right) \exp \left(  j2\pi c \frac{\log \omega}{\log k} \right),
\label{altesOrig}
\end{equation}
which can be verified to satisfy eq.~(\ref{homogeneous}) by choosing 
\begin{equation}
C(n)=k^{n\nu+\frac{n^2}{2}}\exp{\left(-j2\pi nc\right)},\label{Cn}
\end{equation}
The set of real constants $\{A,\nu, k, c\}$ in eq.~(\ref{altesOrig}) arises through grouping of fixed terms, and can be considered as the parameterization of a family of waveforms. As observed by Altes, these waveforms are optimally Doppler tolerant, suitable for radar applications. However, the above parameterization is not straightforward to work with from a wavelet design perspective, which is why, in the following, a more familiar set of filtering parameters is presented.

\begin{figure}[t]
\centering
    \includegraphics[width=8cm,totalheight=3.5cm]{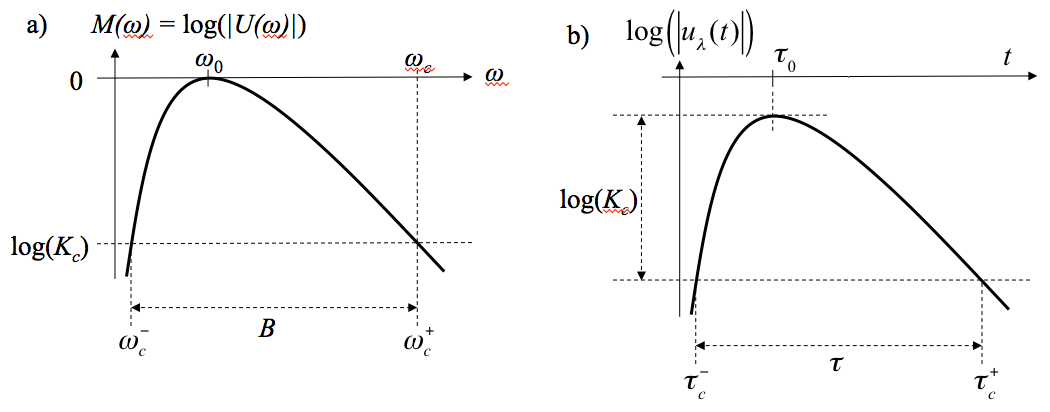}
	\caption{The reparamaeterized Altes chirplet: a) Log-magnitude frequency response, with center frequency $\omega_0$, bandwidth $B$. and upper/lower cutoff frequencies $\omega^\pm_c$. We specify a unit passband response, $|U(\omega_0)|=1$. b) The log-magnitude time-domain envelope, concentrated at $\tau_0$ with delay spread $\tau$. The waveform support is bounded at $\tau^\pm_c$.}
	\label{magfig}
\end{figure}

\subsection{Center Frequency, Cutoff Frequency \& Bandwidth}\label{magnitude}
Eq.~(\ref{altesOrig}) represents a bandpass waveform and should therefore be specific to a center frequency $\omega_0$ and a pair of upper and lower cutoff frequencies $\omega_c^\pm$, equivalent to some bandwidth $B$.  Let the log-magnitude response be defined as
\begin{eqnarray}
M(\omega) &\stackrel{\Delta}{=}& \log\left|U(\omega)\right|\\
&\stackrel{(\ref{altesOrig})}{=}& \log A + \nu \log \omega - \frac{1}{2}\frac{\log^2\omega}{\log{k}}\label{M1}
\end{eqnarray}
and, without loss of generality, be specified with a unit passband response (0~dB gain). This yields a maximum response $M(\omega_0)=0$ at $M'(\omega_0)=0$ from which we get 
\begin{equation}
\omega_0 = k^\nu \label{w0}
\end{equation}
as previously noted in \cite{flandrin1990generalized}, and thereby
\begin{equation}
A=k^{-\nu^2/2}\label{Aeq}.
\end{equation}
As illustrated in Fig.~\ref{magfig}a, the magnitude response $|U(\omega)|$ drops away from unity at $\omega_0$ to some lower level $K_c>0$ at cutoff frequency $\omega_c$ giving, from equations (\ref{M1}), (\ref{w0}) and (\ref{Aeq}) \begin{eqnarray}
M(\omega_c) & = & \nu \log\omega_c - \frac{\nu}{2}\log\omega_0
-\frac{\nu}{2}\frac{\log^2\omega_c}{\log{\omega_0}} \\
&\stackrel{!}{=}&\log(K_c)\;\;\;,\;\;\;0<K_c<1.
\label{Kc0}
\end{eqnarray}
By defining the positive constant
\begin{equation}
\kappa_c  \stackrel{\Delta}{=}-\frac{\log{K_c}}{\log^2\frac{\omega_c}{\omega_0}}>0.
\label{kapc}
\end{equation}
we can easily reparameterize the Altes constants $\nu$ and $k$ as
\begin{eqnarray}
\nu&=&2\kappa_c\log\omega_0,\label{nu}\\
k &=& \exp\left(\frac{1}{2\kappa_c}\right)~.
\label{k}
\end{eqnarray}
Substituting eq.'s~(\ref{nu}), (\ref{k}) \& (\ref{Aeq}) in eq.~(\ref{M1}) gives the much simplified magnitude response expression
\begin{equation}
M(\omega)=-\kappa_c\log^2\frac{\omega}{\omega_0}.
\label{M2}
\end{equation}
The definition of $\kappa_c$ allows us to write
\begin{equation}
\omega^\pm_c \stackrel{(\ref{kapc})}{=} \omega_0 \exp\left(\pm\sqrt{\frac{-\log{K_c}}{\kappa_c}}\right) \textrm{rad/s}.\label{wc}
\end{equation}
We can specify the upper and lower cutoff frequencies by respectively choosing $\omega_c$ greater or less than center frequency $\omega_0$. We recommend to disambiguate in favor of the upper one, and speak of \emph{the} cutoff frequency $\omega_c\stackrel{\Delta}{=}\omega_c^+>\omega_0$. The upper cutoff exerts control over decay of $U(\omega)$ at higher frequencies, and hence waveform regularity (Section~\ref{regularity}). There is less need to worry about the lower cutoff because of the zero in $U(\omega)$ at $\omega=0$ and its extremely rapid decay at low frequencies. Accordingly, the bandwidth is defined simply
\begin{eqnarray}
B &\stackrel{\Delta}{=}& \omega_c^+ - \omega_c^- \label{B}\\
&\stackrel{(\ref{wc})}{=}&\omega_0\left(\frac{\omega_c}{\omega_0}-\frac{\omega_0}{\omega_c}
\right)\;\textrm{rad/s}\label{BW}
\end{eqnarray}
with the lower cutoff frequency given by $\omega_c^- = \omega_0^2/\omega_c$.

For the presented formulation to be useful in practice, a value must be placed on $K_c = |U(\omega_c)|$ used in the definition of  cutoff frequency in eq.~(\ref{Kc0}). For the twin objectives of wavelet frequency localization and avoidance of discrete-time aliasing elaborated in Section~\ref{paramSelec}, it makes sense to place tight restrictions on $K_c$. We use a level corresponding to $-40$dB throughout, which of course can be varied depending on application requirements. To be pedantic, this level represents 
\begin{equation}
20\log_{10}\left|U(\omega_c)\right| = -40\;\textrm{dB}
\end{equation}
\begin{equation}
\Rightarrow K_c = 10^{-\frac{40}{20}} = 0.01.\label{Kc}
\end{equation}

\subsection{Chirp Rate $\lambda$ and Re-parameterization of the Altes Wavelet}
\label{chirpRate}
As foreseen in Section~\ref{constQsection}, the phase and magnitude response behaviours of the Altes' waveform in eq.~(\ref{altesOrig}) are both governed by single parameter $k$. In order to decouple them, and allow for simpler chirplet design, we introduce 
\begin{equation}
\log \lambda \stackrel{\Delta}{=} \frac{\log k}{c}.\label{c}
\end{equation} 
The phase response of $U(\omega)$ then corresponds to hyperbolic chirping in the time domain, with $\lambda$ controlling the chirp rate. 
The time-domain pulse given by inverse Fourier transform 
of eq.~(\ref{altesOrig})
\begin{equation}
u(t) = \mathcal{F}^{-1}\left\{U(\omega)\right\}\label{ut}
\end{equation}
is shown in \cite{altes1980models} to have phase response 
\begin{equation}
\phi(t) \propto \log t,
\end{equation}
a logarithmic function of time. Instantaneous frequency is
\begin{equation}
\omega_I(t)  \stackrel{\Delta}{=}   \frac{\textrm{d}\phi(t)}{\textrm{d}t}
\propto\frac{1}{t} 
\end{equation}
which describes a hyperbola in the time-frequency plane. Since the instantaneous period $T_I(t) =  2\pi/\omega_I(t) \propto t$  is linear in time, signals of type $u(t)$ are interchangeably (and correctly) referred to as having linear period modulation (LPM), hyperbolic frequency modulation (HFM) or logarithmic phase modulation \cite{kroszczynski1969pulse}, or as being log-periodic, which is preferred in the financial and physics literature \cite{geraskin2013everything}. 

The proposed re-parameterization now emerges from eq.~(\ref{altesOrig}) through equations (\ref{w0}), (\ref{Kc0}) and (\ref{c}),  as 
\begin{equation}
U(\omega) = 
    \exp\left(\log{K_c} {\log^2\frac{w}{w_0} \over \log^2\frac{\omega_c}{\omega_0}}\right)\exp\left(j2\pi\frac{\log\omega}{\log\lambda}\right),\;\;\;\forall\;\omega>0.\label{newAltes2}
\end{equation}
with $\omega_0>0$ and $\lambda>0$, or more concisely from eq.~(\ref{kapc})
\begin{equation}
U(\omega) = 
    \exp\left(-\kappa_c \log^2\frac{w}{w_0}\right)\exp\left(j2\pi\frac{\log\omega}{\log\lambda}\right),\;\;\;\forall\;\omega>0.\label{newAltes}
\end{equation}

\begin{figure}[t]
\centering
    \includegraphics[width=8.5cm, totalheight = 6.5cm]{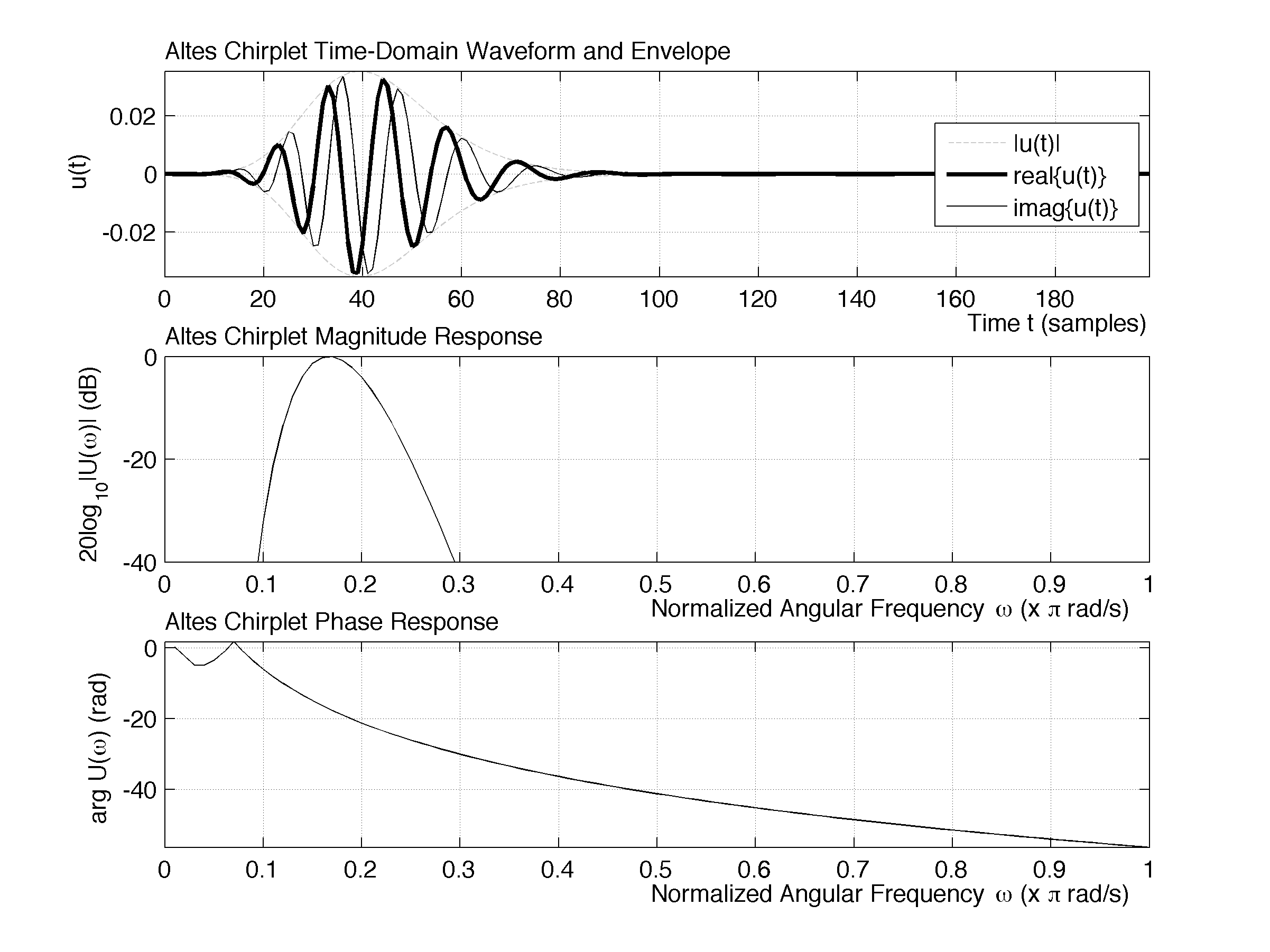}
	\caption{Example 2 demonstrates tunability of familiar wavelet parameters such as center frequency $\omega_0$, bandwidth $B$ and chirp rate $\lambda$, as a result of our re-parameterization. In this case $\{\omega_0,B,\lambda\} = \{\frac{\pi}{6},\frac{\pi}{5},\frac{3}{4}\}$, and a unit sampling interval is chosen.}
	\label{example2}
\end{figure}

This parameterisation makes clear that $\lambda$ is the scaling ratio of the discrete scale invariance symmetry. It is the ratio of the local periods of the successive oscillations in the chirp. $\omega_0$ is the center frequency for which $|U(\omega)|$ is maximum (here normalised to $1$). $\kappa_c$ controls the lower and upper cut-off frequencies $\omega_c^\pm$ in eq.~(\ref{wc})
beyond which the signal amplitude falls off by more than a certain pre-specified level $K_c$ from eq.~(\ref{Kc}).
For a given $K_c$, the Altes chirplet can thus be parameterised by the triplet $\{\omega_0, \kappa_c, \lambda\}$.
Alternatively, $\kappa_c$ can be replaced by the bandwith $B$ from eq.~(\ref{BW})
(see figure \ref{example2}) or cut-off frequency $\omega_c$ from eq.~(\ref{wc}) as in eq.~(\ref{newAltes2}) (see figures \ref{example3} and \ref{example4}).

As the chirp rate parameter $\lambda\rightarrow 1$ in eq.~(\ref{newAltes}), the oscillations become faster and faster until 
a singularity occurs at $\lambda = 1$. To avoid it, we must consider $0<\lambda<1$ or $\lambda>1$. Note also, however, that replacing $\lambda$ by $\frac{1}{\lambda}$ simply changes the sign of the complex exponential in eq.~(\ref{newAltes}), equivalent to a frequency domain conjugation. This in turn represents time reversal and conjugation of the chirplet in the time domain, an example of which will be presented in Fig.~\ref{example4}. Making the dependence on chirp rate explicit by subscript, we can write
\begin{equation}
u_\frac{1}{\lambda}(t) = u^*_\lambda(-t).
\end{equation}
Later, in considering the effect of $\lambda$ on time-frequency localization, we thus need only examine $\lambda\in[0,1)$. The effects for $\lambda\in(1,\infty)$ are identical, and found by reciprocation of $\lambda$.

In contrast to the original Fourier domain specification of $U(\omega)$ \cite{altes1975n2}, we propose to zero the non-positive frequencies and define  
\begin{equation}
U(\omega)\stackrel{\Delta}{=}0,\;\;\;\forall \; \omega\le0 \label{analytic2}
\end{equation} 
While Altes prefered to impose Hermitian symmetry on $U(\omega)$ to ensure a real wavelet, preservation of the analytic form retains useful phase and envelope properties --- in particular, this allows natural definitions for instantaneous amplitude and instantaneous frequency of arbitrary signals being analyzed\cite{flandrin2001time}. From (\ref{analytic2}) and (\ref{newAltes}), $U(\omega)$ is continuous at $\omega=0$ since  
\begin{equation}
\lim_{\omega\rightarrow0^+}U(\omega) = \lim_{\omega\rightarrow0^-}U(\omega) = U(0) = 0.\label{analytic0}
\end{equation}
This Altes chirp $U(\omega)$ is bandpass and frequency localized, a minimum requirement for formal admissibility  as a wavelet. 

\begin{figure}[t]
\centering
    \includegraphics[width=8.5cm, totalheight = 6.5cm]{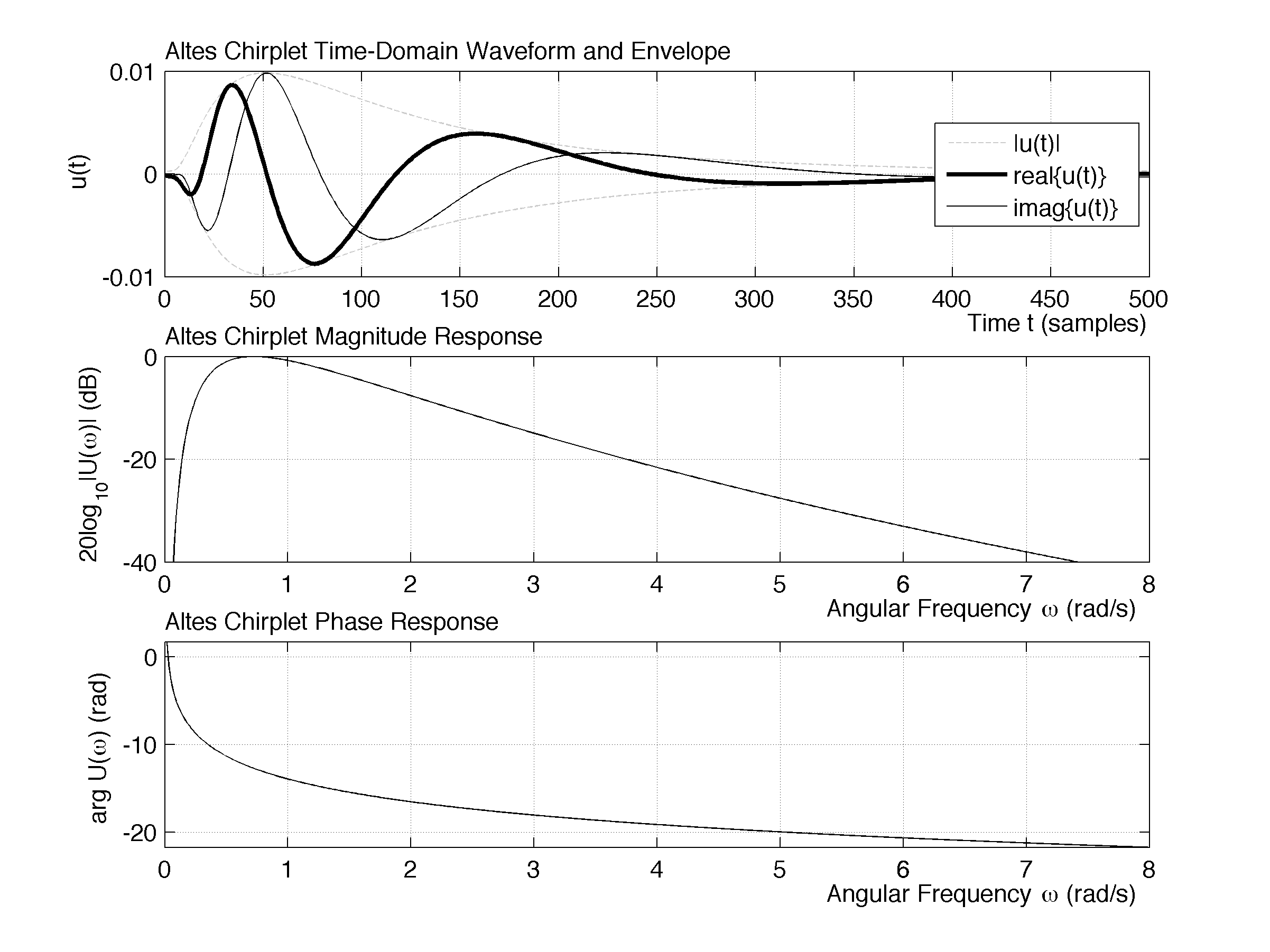}
	\caption{Example 1 is taken from the original Altes paper \cite{altes1975n2} using his parameter set $\{\nu,k,c\}=\{-0.55,1.8,-0.35\}$. There is no frequency normalization, and  it is not obvious how the parameters relate to the observed waveform. We find $\{\omega_0,B,\lambda\} = \{0.7328,7.344,0.1865\}$.}
	\label{example1}
\end{figure}

\subsection{Examples}
\label{examples}
Four examples are now presented to illustrate the behavior of the Altes waveform as a function of its parameters. In each case, we show the analytic time-domain waveform $u(t)$ from eq.~(\ref{ut}), as well as the magnitude and phase responses of $U(\omega)$ in the frequency domain from eq.~(\ref{newAltes}). 

In the first example, Fig.~\ref{example1} recreates a waveform from Altes' original paper with $\{\nu,k,c\}=\{-0.55,1.8,-0.35\}$. A primary difference to the original exposition is that our chirp is complex, and both the imaginary component and complex envelope can be observed in addition to the real waveform. As trivia, we can evaluate from eq.~(\ref{w0}) that this bandpass waveform has center frequency $\omega_0=0.7328$~rad/s, as well as providing from equations~(\ref{c}), (\ref{wc}) and (\ref{BW}), 
 the previously  unavailable chirp rate $\lambda=0.1865$, cutoff frequency $\omega_c=7.4146$~rad/s,  and bandwidth $B=7.3440$~rad/s, assuming as we have a -40dB cutoff, i.e. $K_c=10^{-2}$.

\begin{figure}[t]
\centering
    \includegraphics[width=8.5cm, totalheight = 6.5cm]{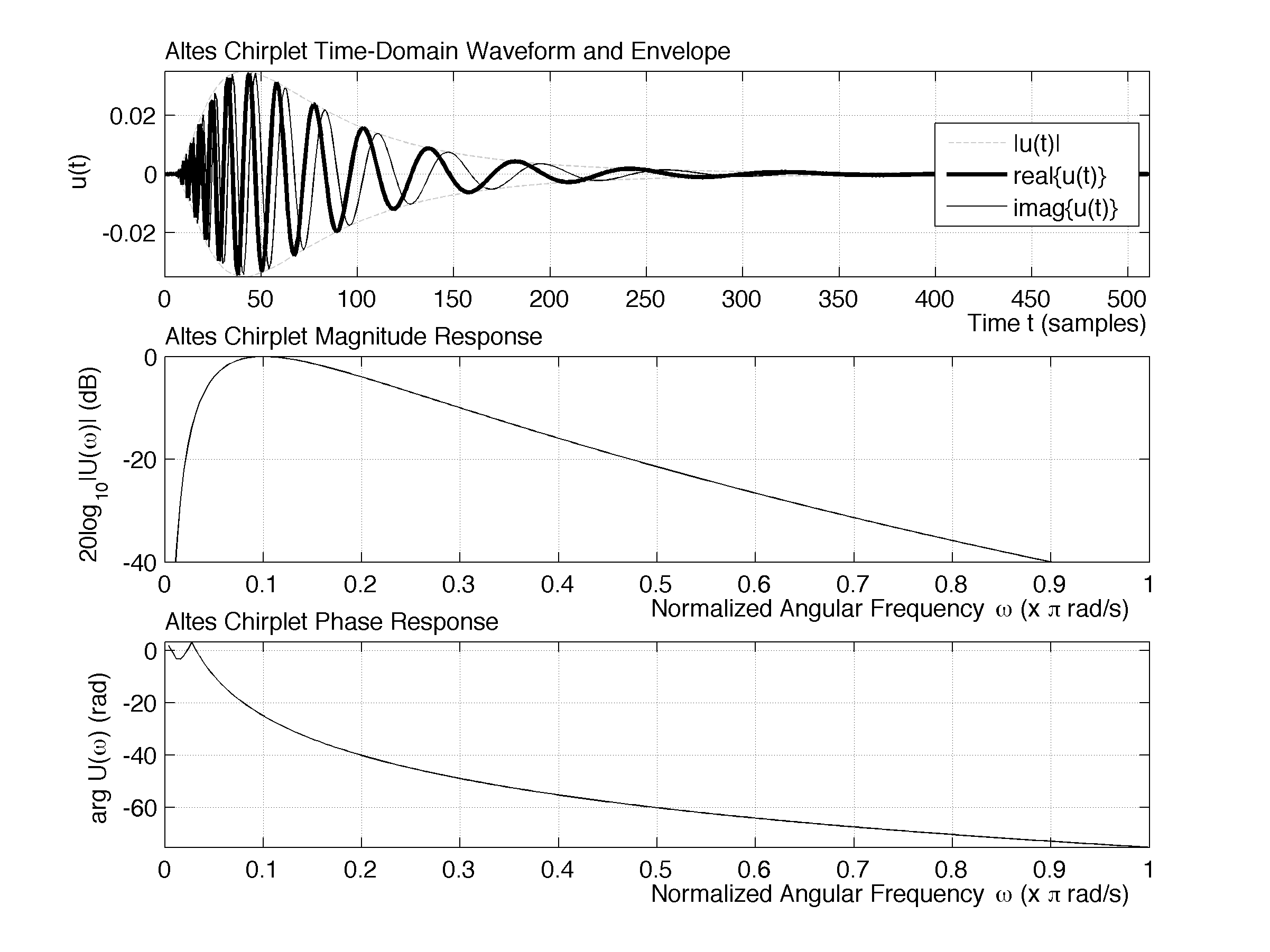}
	\caption{In example 3, the chirplet is tuned via its cutoff frequency rather than its bandwidth: $\{\omega_0,\omega_c,\lambda\} = \{\frac{\pi}{10},\frac{9\pi}{10},\frac{3}{4}\}$. This allows tight control of magnitude response decay at high frequencies, which influences regularity (wavelet smoothness) and aliasing in a discrete time implementation.}
	\label{example3}
\end{figure}
\begin{figure}[t]
\centering
    \includegraphics[width=8.5cm, totalheight = 6.5cm]{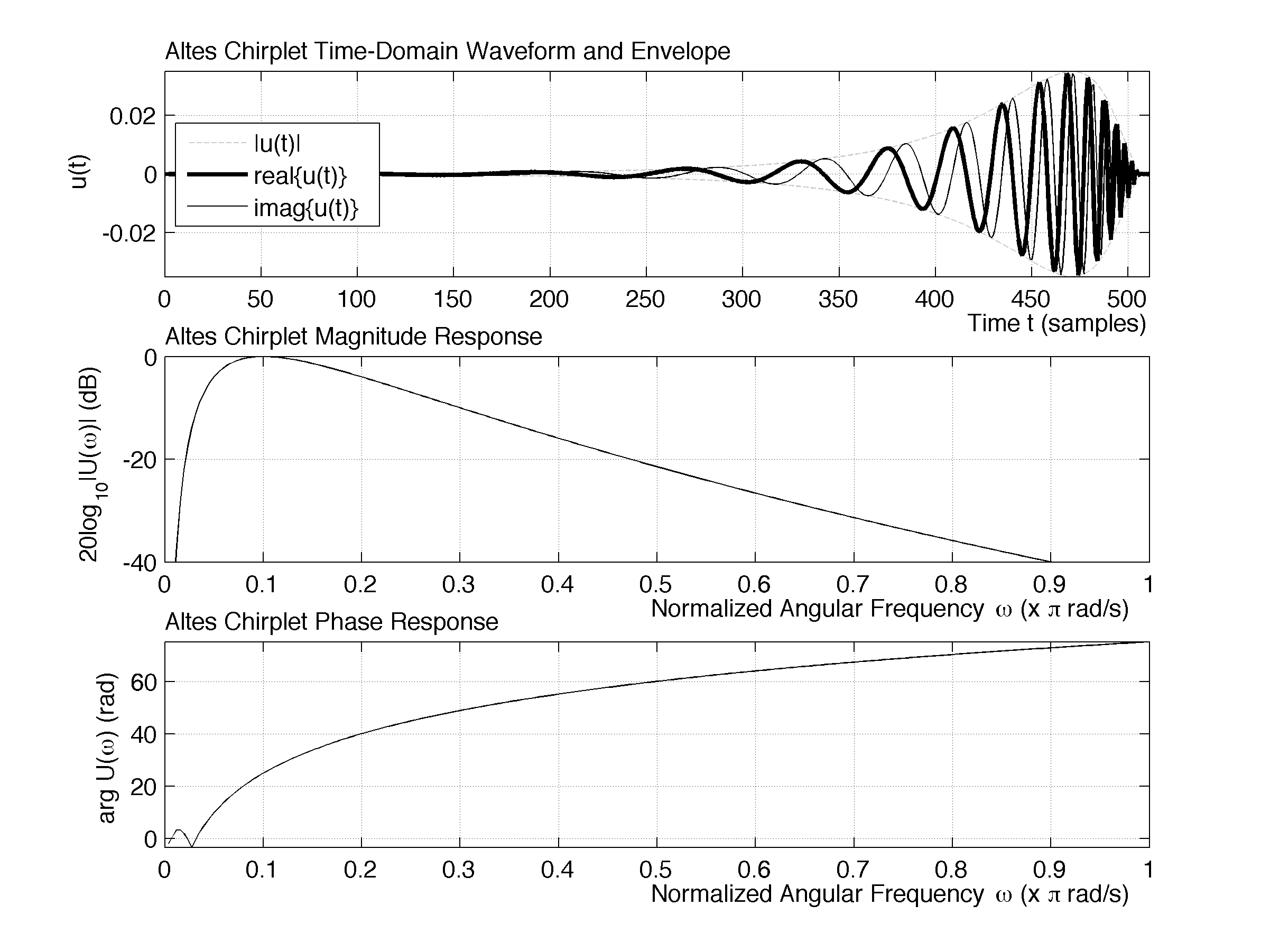}
	\caption{Example 4 is chosen to show how a reciprocal chirp-rate $\lambda$ affects the phase- of the chirplet in the time-domain. We have used $\{\omega_0,\omega_c,\lambda\} = \{\frac{\pi}{10},\frac{9\pi}{10},\frac{4}{3}\}$. Compared with example 3, This results in time-reversal and conjugation, as explained in Section~\ref{chirpRate}.}
	\label{example4}
\end{figure}

In the second example, we choose to show in Fig.~\ref{example2} how our parameterization allows practical specification of center-frequency, bandwidth and chirp rate, in this case $\{\omega_0,B,\lambda\} = \{\frac{\pi}{6},\frac{\pi}{5},\frac{3}{4}\}$. In the graphic, it can be seen that the $-40$~dB bandwidth $B$ is $0.2\pi$ as specified. When implementing in discrete time, it is useful, although not necessary, to work in units of normalized frequency, such that the sampling frequency $\omega_s=2\pi$ corresponds to a unit sampling interval. This convention is adopted in the remainder  which fixes the Nyquist rate at $\omega_\textrm{Nyq} = \pi$ for the coming discussion on discrete-time implementation.

The third example illustrates the tuning of waveform cutoff frequency using parameters $\{\omega_0,\omega_c,\lambda\} = \{\frac{\pi}{10},\frac{9\pi}{10},\frac{3}{4}\}$. In Fig.~\ref{example3}, it can be seen that the -40~dB cutoff is tightly tuned to $\omega_c=\frac{9\pi}{10}$ as specified. The wider bandwidth opens a greater range of frequencies over which to chirp. This is visible in the waveform, when compared with Fig.~\ref{example2}.

Our final example is chosen to show the effect of the chirp rate parameter $\lambda$. From eq.~(\ref{newAltes}), it is clear that $0<\lambda<1$ results in negative phase response, while $\lambda>1$ yields positive phase response. We showed that replacing $\lambda$ by $\frac{1}{\lambda}$ amounts to conjugation and time reversal of the Altes chirp. The effect is clear to see in Fig.~\ref{example4} in which the Altes chirplet is parameterized identically to example 3, with the exception of the chirp rate, which is inverted: $\{\omega_0,\omega_c,\lambda\} = \{\frac{\pi}{10},\frac{9\pi}{10},\frac{4}{3}\}$. The effects of chirp-rate on wavelet delay-spread will be examined further in Section~\ref{tfloc} on time-frequency localization.

\section{The Hyperbolic Altes Chirplet Transform}\label{wavelet}
Up to this point, we have arrived at the description of a family of \emph{waveforms} reparameterized in eq.~(\ref{newAltes}) from Altes' original frequency domain representation in eq.~(\ref{altesOrig}). In this Section, we show that we are actually dealing with a family of \emph{wavelets} in the formal sense, with desirable additional properties.


\subsection{Admissibility}
The continuous wavelet transform (CWT) of a real or complex, square integrable signal $s(t)$ at position $b$ and scale $a$ is 
\begin{equation}
C_\psi(a,b) = \int_\mathbb{R}s(t)\frac{1}{\sqrt{a}}\psi^*\left(\frac{t-b}{a}\right)\,\textrm{d}t,\;\;\;a\in\mathbb{R}^+,\;b\in\mathbb{R}
\label{cwt}
\end{equation}
where $\psi(t)$ is said to be an \emph{admissible} wavelet if it satisfies the fairly loose conditions that it is square integrable and sufficiently band-limited \cite{aguiar2014continuous}. Calder\'{o}n's reproducing identity tells us that in this case the original signal $s(t)$ can be recovered exactly from its wavelet coefficients $C_\psi(a,b)$ by the inverse transform 
\begin{equation}
s(t) = \frac{1}{\mathcal{C}_\Psi}\int_\mathbb{R}\int_{\mathbb{R}^+} C_\psi(a,b)\frac{1}{\sqrt{a}}\psi\left(\frac{t-b}{a}\right)
\,\textrm{d}a\,\textrm{d}b,
\end{equation}where $\mathcal{C}_\Psi$ is the admissibility constant. The formal admissibility conditions on $\psi(t)$ are
\begin{eqnarray}
\mathcal{E}_\psi&\stackrel{\Delta}{=}&\int_\mathbb{R}\left| \psi(t) \right|^2\,\textrm{d}t<\infty\label{Csqi}\\
\mathcal{C}_\Psi&\stackrel{\Delta}{=}&\int_\mathbb{R}\frac{\left| \Psi(\omega)\right|^2}{\left| \omega\right|}<\infty
\label{Cadm}
\end{eqnarray}
corresponding to square-integrability and transform invertability, respectively. $\mathcal{E}_\psi$ is the wavelet energy and we have used the Fourier Transform $\mathcal{F}\left\{\psi(t)\right\}=\Psi(\omega)$.

The CWT in eq.~(\ref{cwt}) represents the correlation of $s(t)$ with scaled and shifted versions of the wavelet
\begin{equation}
\psi_{a,b}(t) = \frac{1}{\sqrt{a}}\psi\left(\frac{t-b}{a}\right). \label{psiab}
\end{equation}
The factor $\frac{1}{\sqrt{a}}$ ensures that $\Vert\psi_{a,b}(t)\Vert$ is independent of $\{a,b\}$, and often the mother wavelet $\psi(t)$ is normalized so that 
\begin{equation}
\mathcal{E}_\psi=\Vert\psi(t)\Vert_2=\Vert\psi_{a,b}(t)\Vert_2=1.\label{norm}
\end{equation}
The admissibility conditions tell us that $\psi(t)$ is a finite energy pulse with a frequency response that decays at high frequencies, and which must have no DC component. In simpler words, it is a frequency localized bandpass waveform.

\newtheorem{proposition}{Proposition}
\begin{proposition}
The Altes waveform $u(t)=\mathcal{F}^{-1}\left\{U(w)\right\}$ is an admissible wavelet.
\end{proposition}
{\bf Proof}. A log-normal random variable $X$ with $\log(X)\sim\mathcal{N}(\mu,\sigma^2)$ has, by definition, probability density function and expected value given by 
\begin{eqnarray}
p_X(x) &=& \frac{1}{x\sigma\sqrt{2\pi}}
\exp\left(
-\frac{\left(\log x-\mu\right)^2}
{2\sigma^2}
\right),x>0
\\
E[X]&=&\exp\left(\mu+\frac{\sigma^2}{2}\right).\label{meanLogNorm}
\end{eqnarray}
The mean is defined by 
\begin{equation}
E[X]\stackrel{\Delta}{=}\int_{-\infty}^\infty xp_X(x)\,\textrm{d}x
\end{equation}
so the equality (\ref{meanLogNorm}) translates to
\begin{equation}
\frac{1}{\sigma\sqrt{2\pi}}\int_0^\infty\exp\left(
-\frac{\left(\log x-\mu\right)^2}
{2\sigma^2}
\right)
\,\textrm{d}x
=\exp\left(\mu+\frac{\sigma^2}{2}\right).
\end{equation}
Replacement $x\rightarrow\omega$, $\mu\rightarrow\log\omega_0$ and  $\sigma^2\rightarrow\frac{1}{4\kappa_c}$ gives
\begin{equation}
\sqrt{\frac{2\kappa_c}{\pi}}\int_0^\infty 
e^{
-2\kappa_c\log^2\frac{\omega}{\omega_0}
}\,\textrm{d}\omega
=e^{{\log\omega_0+{\frac{1}{8\kappa_c}}}}
\end{equation}
and so from eq.~(\ref{newAltes})
\begin{equation}
\frac{1}{2\pi}
\int_{-\infty}^\infty\left| U(\omega)\right|^2\,\textrm{d}\omega
=\frac{\omega_0}{\sqrt{8\pi\kappa_c}}\exp\left({\frac{1}{8\kappa_c}}\right)
\label{sqi2}
\end{equation}
From Parseval's Theorem
\begin{equation}
\int_{-\infty}^\infty\left| u(t)\right|^2\,\textrm{d}t=
\frac{1}{2\pi}
\int_{-\infty}^\infty\left| U(\omega)\right|^2\,\textrm{d}\omega
\stackrel{(\ref{sqi2})}{<}\infty\label{parseval}
\end{equation}
proving square integrability by eq.~(\ref{Csqi}).  As an aside, equations~(\ref{sqi2}) and (\ref{parseval}) can be used for normalization in eq.~(\ref{norm}).

By making the substitution $x\leftarrow\log\frac{\omega}{\omega_0}$ in eq.~(\ref{Cadm}) we get
\begin{eqnarray}
\mathcal{C}_U &\stackrel{(\ref{newAltes})}{=} &\int_0^\infty\exp^2\left( 
-\kappa_c\log^2\frac{\omega}{\omega_0}
\right) \,\frac{\textrm{d}\omega}{\omega}\\
&=&\int_{-\infty}^\infty\exp\left(-2\kappa_cx^2\right)\,\textrm{d}x\\
&=&\sqrt{\frac{\pi}{2\kappa_c}}\\
&<&\infty
\end{eqnarray}
proving invertability and, thereby, admissibility. 
Not only can the Altes waveform now be correctly called a wavelet, but since it is a log-periodic chirp, we can also refer to it to as the Altes chirplet, and its application within a CWT as the Hyperbolic Chirplet Transform, consistent with prior taxonomy. Wavelets that only have the property of admissibility are known as \emph{crude} wavelets, because admissibility is a weak condition and does not guarantee usefulness for signal processing. This leads us to examine other wavelet properties which can enhance their applicability.

\subsection{Regularity}\label{regularity}
Regularity describes the smoothness of a wavelet $\psi(t)$ in the time domain. The order of regularity corresponds to the number of times $\psi(t)$ is continuously differentiable. We say that $\psi(t)$ is bounded and has uniform Lipschitz regularity of order $\alpha>0$ over $\mathbb{R}$ if its frequency weighted magnitude response is Lebesgue integrable according to
\begin{equation}
\int_\mathbb{R}\left|\Psi(\omega)\right|
\left(1+|\omega|^\alpha\right)\,\textrm{d}\omega<\infty.\label{regular}
\end{equation}
This captures the fact that smoothness in time is directly related to the rate of decay at high frequencies. Intuitively, signals with higher frequency content vary more rapidly, and are therefore less regular. 
\begin{proposition}
The Altes chirplet $u(t)$ has infinite regularity. 
\end{proposition}
{\bf Proof}. 
Using $U(\omega)$ from eq.~(\ref{newAltes}), we define 
\begin{eqnarray}
I_0 &\stackrel{(\ref{regular})}{=}& \int_0^\infty\exp\left( 
-\kappa_c\log^2\frac{\omega}{\omega_0}
\right) \,\textrm{d}\omega\nonumber\\&&+
\int_0^\infty\omega^\alpha\exp\left( 
-\kappa_c\log^2\frac{\omega}{\omega_0}
\right) \,\textrm{d}\omega\\
&\stackrel{\Delta}{=}&I_1+I_2.
\end{eqnarray}
Replacing $x\leftarrow\log\frac{\omega}{\omega_0}$ gives
\begin{eqnarray}
I_2 &=& \omega_0^{\alpha+1}\int_{-\infty}^\infty\exp\left( 
-\left(\kappa_cx^2-(1+\alpha)x\right)
\right) \,\textrm{d}x\\&=&
\omega_0^{\alpha+1}\sqrt{\frac{\pi}{\kappa_c}}\exp\left( 
\frac{(1+\alpha)^2}{4\kappa_c}\right)
\end{eqnarray}
The limit $\alpha\rightarrow0$ gives $I_1$ allowing us to compute the definite integral
\begin{eqnarray}
I_0&=&
\omega_0\sqrt{\frac{\pi}{\kappa_c}}e^{
\frac{1}{4\kappa_c}}
\left[
1+\omega_0^\alpha e^{ 
\frac{\alpha^2+2\alpha}{4\kappa_c}}
\right]<\infty
\end{eqnarray}
for all finite $\alpha>0$.
Wavelets are often used in compression applications, whereby a signal is represented by a truncated set of its CWT coefficients. Regular wavelets have the advantage that, when used in a CWT for such application, the undesirable artifacts arising from truncation (e.g. audio distortion) are less noticeable when compared to those produced using less smooth wavelets, even if the compression error magnitude is similar. Infinitely regular wavelets are thus suitable for use in information-coding. The Altes chirplet joins an august family including the Morlet, Mexican Hat, Meyer, Gauss and Shannon wavelets, all infinitely continuously differentiable in time.

\subsection{Vanishing Moments}\label{vanmom}
If wavelet $\psi(t)$ has $M$ vanishing moments, then it is orthogonal to all polynomials of order $M-1$. A richer set of signals can be represented with a sparser set of coefficients when the mother wavelet of the analyzing CWT has higher $M$. The number of vanishing moments is the highest integer $M$ such that in the time domain
\begin{equation}
\int_{-\infty}^\infty t^m\psi(t)\,\textrm{d}t=0\;\;\forall\;\;m\in\{0,1,2,\ldots,M\}
\end{equation}
or equivalently in the frequency domain
\begin{equation}
\left.\frac{\textrm{d}^m\Psi(\omega)}{\textrm{d}\omega^m}\right|_{\omega=0}=0\;\;\forall\;\;m\in\{0,1,2,\ldots,M\}.\label{vmf}
\end{equation}
The latter version makes it clear that vanishing moments flatten the wavelet response $\Psi(\omega)$ around DC. However, steeper response decay at low frequencies narrows the wavelet bandwidth from below, and there must be a corresponding time dilation. Thus, a higher number of vanishing moments comes at the cost of increasing support in the time domain i.e. longer wavelets. This will be addressed in Section~\ref{paramSelec}.
\begin{proposition}
The Altes chirplet has an infinite number of vanishing moments.  
\end{proposition}
{\bf Proof}. 
Since $U(\omega)=0,\omega\le0$ it suffices to show that 
\begin{equation}
\lim_{\omega\rightarrow0^+}\frac{\textrm{d}^nU(\omega)}{\textrm{d}\omega^n}=0\;\;,\;\;\forall\;\;n\in\mathbb{Z}^+.
\end{equation}
It is obvious that the exponential form of the Altes wavelet $U(\omega)$ from eq.~(\ref{newAltes})
renders it infinitely continuously differentiable over the positive frequencies $\omega>0$. While computing these derivatives is cumbersome, by rewriting 
\begin{equation}
U(\omega)=\exp(Q(\omega))\;\;,\;\;\omega >0\label{expAltes}
\end{equation}
it can be readily found from chain and product rules that they take the form
\begin{equation}
\frac{\textrm{d}^nU(\omega)}{\textrm{d}\omega^n} = U(\omega)\sum_{p=1}^n\sum_{q=1}^p
a_pQ_p^{b_p}(\omega)
a_qQ_q^{b_q}(\omega)\label{dnUdwn}
\end{equation}
where 
\begin{eqnarray}
Q_i(\omega) &=& \frac{\textrm{d}^iQ(\omega)}{\textrm{d}\omega^i}\\
&\stackrel{(\ref{newAltes})(\ref{expAltes})}{=}&\frac{c_{i,0}\log\omega+c_{i,1}+jc_{i,2}}{\omega^i}.
\end{eqnarray}
There are a finite number of terms in the double sum of eq.~(\ref{dnUdwn}). This number is independent of $\omega$, as are the integer constants $a_i, b_i$ and the real constants $c_{i,k}$. These coefficients must be found by computation, but we do not need them to observe that, as $\omega$ shrinks to zero  in eq.~(\ref{dnUdwn}),  the magnitude of $U(\omega)$ shrinks far more quickly than the products of $Q_i^{b_i}(\omega)$ explode. The former has order of growth  $\exp(-\log^2\omega)$ which dominates the latter, whose order of growth is only $(\log\omega)/\omega^n$ at the origin. This dominance holds true for any $n\in\mathbb{Z}^+$.

\subsection{Scale Invariance}\label{scalInv}
When the Altes Chirplet is used as the mother wavelet in the CWT, we call this the Hyperbolic Chirplet Transform (HCT). The self-similarity of the Altes Chirplet, as it emerges from the homogeneity of eq.~(\ref{homogeneous}), leads to the extraordinary property of transform scale invariance, ie. the HCT can be computed trivially at any scale, from knowledge of one scale only. 
\begin{proposition}
The HCT is scale invariant.
\end{proposition}
{\bf Proof}. 
The continuous wavelet transform of eq.~(\ref{cwt}) can also be defined in the frequency domain as 
\begin{equation}
C_\psi(a,b)=\frac{\sqrt{a}}{2\pi}\int_\mathbb{R}S(\omega)\Psi^*(a\omega)e^{j\omega b}\textrm{d}\omega.
\label{cwtFreq}
\end{equation}
Taking $n=1$ in eq.'s~(\ref{homogeneous}) and (\ref{Cn}) and using eq.~(\ref{c}) gives
\begin{equation}
\omega U(\omega) = k^{\nu+\frac{1}{2}}\exp\left(
-j2\pi\frac{\log k}{\log \lambda}
\right)U\left(\frac{\omega}{k}\right).\label{homog2}
\end{equation}
For $k\rightarrow \frac{1}{a}$ in eq.~(\ref{homog2}) with $U(\omega)\rightarrow\Psi(\omega)$ in eq.~(\ref{cwtFreq})
and $U(\omega)$ given by (\ref{newAltes}), we get
\begin{equation}
C_u(a,b) = \frac{g(a)}{2\pi}\int_\mathbb{R}S(\omega)
\omega U^*(\omega )e^{j\omega b}\textrm{d}\omega\label{hct}
\end{equation}
with
\begin{equation}
g(a) \stackrel{\Delta}{=} a^{1+2\kappa_c \log\omega_0}\exp\left(j2\pi\frac{\log a}{\log \lambda}\right)\label{glam}
\end{equation}
i.e. independent of $\omega$. We have used eq.~(\ref{nu}) to replace $\nu$ in line with the new parameterization. From the above, it can readily be shown that, for any real multiplier $m>0$, we have
\begin{equation}
C_u(ma,b) \stackrel{(\ref{hct})(\ref{glam})}{=} g(m)C_u(a,b)\label{wsi}
\end{equation}
with $g(m)$ independent of $a$ and $b$, proving scale invariance. 
The integrand in eq.~(\ref{hct}) is independent of scale $a$ which means that the Hyperbolic Chirplet Transform $C_u(a,b)$ can be computed once at scale $a$, and the coefficients $C_u(ma,b)$ can be found at all other wavelet scales $ma$ via eq.~(\ref{wsi}). This results from the self-similarity of the Altes wavelet that performs in a sense a full multiscale analysis with just one scale of magnification $a$. Of course the chirplet transform values must be re-computed for different shift values $b$.

Note that, at unit scale, $a=1$ and $g(1)=1$. In this case eq.~(\ref{hct}) has the time domain equivalent
\begin{equation}
C_u(1,b) \stackrel{(\ref{cwt})(\ref{psiab})}{=} s(t) \star u(t-b),
\end{equation}
where $\star$ is cross-correlation. In words, we find the CWT coefficients at unit scale by measuring the correlation of signal $s(t)$ with the delayed Altes chirplet, and use eq.~(\ref{wsi}) to find the coefficients at other scales. In practice, the HCT is implemented digitally, with the waveform being discretized in both time and amplitude. As such a full, infinite-support, infinite-scale, multiresolution analysis will not be achievable from eq.~(\ref{wsi}), but will depend on the specific implementation. Nevertheless, the HCT can be implemented as an extremely efficient transform as a result of the self-similarity of the Altes chirplet. 

\section{Parameter Selection}\label{paramSelec}
The Altes chirplet was shown in Section~\ref{vanmom} to have an infinite number of vanishing moments. This requires infinite support in the time domain, suggesting very long wavelets. In this section, it is seen that, through suitable parameterization, the chirplet's time-decay can be tailored for desired localization, in trade-off with its frequency domain bandwidth and chirping flexibility. In addition, the CWT, when implemented in software or hardware, must be approximated in discrete-time with a finite number of samples using appropriate time-frequency sampling. The influence of discrete-time implementation on parameter selection is therefore also discussed. As there is no trivial time domain representation of the Altes chirplet from eq.~(\ref{ut}), this part of the study is numerical.

\subsection{Delay Spread}\label{dsp}
We start by defining the delay-spread of the waveform in a manner  similar to how its bandwidth is specified in the frequency domain in Section~\ref{magnitude}. Time localization implies that the time domain envelope $|u_\lambda(t)|$ rises gradually from zero to some peak at $t=\tau_0$ where the waveform energy is concentrated, before falling away again to zero, as in Figures ~\ref{example1}--\ref{example4} for example, and illustrated clearly in Fig.~\ref{magfig}b. We define the delay spread empirically as the time $\tau$ between the earliest appearance of significant energy at $t=\tau_c^-$, and the later time $t=\tau_c^+$ beyond which the energy appears to have vanished: 
\begin{equation}
\tau=\tau_c^+-\tau_c^-.
\end{equation}
For this to be precisely specified, some threshold amplitude is needed, which defines the presence or absence of wave energy at cutoff points $\tau_c^\pm$. We use the same -40~dB level (\ref{Kc}) used in defining the waveform bandwidth in Section.~\ref{magnitude}. Normalizing to the peak magnitude of the time domain waveform, we define this amplitude threshold as 
\begin{eqnarray}
\frac{\left|u_\lambda(\tau_c^\pm)\right|}{\left|u_\lambda(\tau_0)\right|} \stackrel{!}{=} K_c = 0.01.
\end{eqnarray}
While this is not the most conventional definition of delay spread, it provides consistency with eq.~(\ref{Kc}) for time-frequency localization analysis, and is certainly valid.

The questions can now be asked: how does the Altes wavelet delay-spread $\tau$ depend on the parameterization $\{\omega_0,\omega_c,\lambda\}$ from eq.~(\ref{newAltes2}), and how does it trade off against the chirplet bandwidth? Indeed, what are useful values of, or limits on, these parameters for signal processing applications? 

\subsection{The Efficient Frontier of Delay Spread vs. Bandwidth} \label{tfloc}
\begin{figure}[t]
\centering
    \includegraphics[width=8cm, totalheight=5.5cm,angle=0]{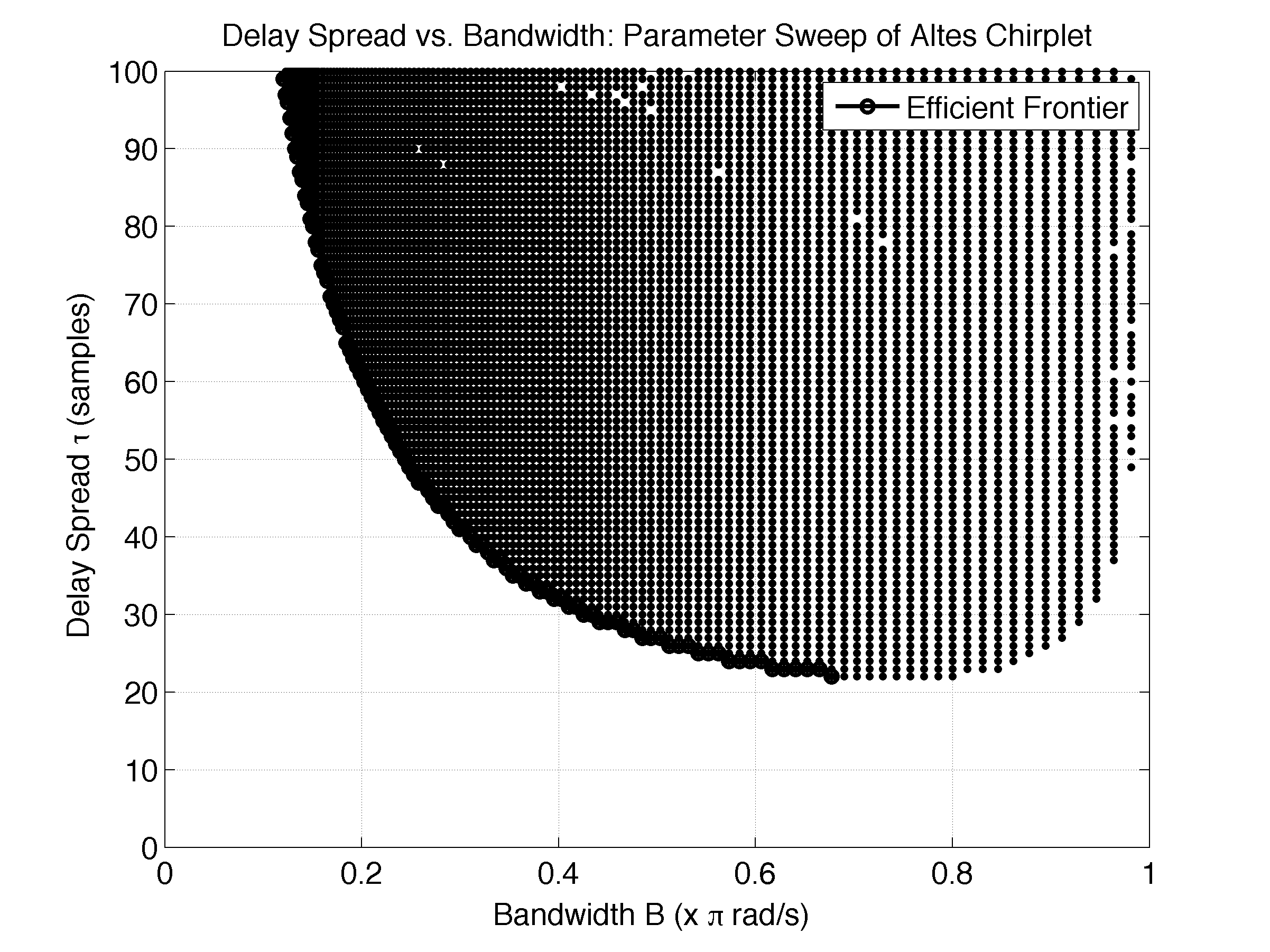}
	\caption{Each dot in the figure represents one instance of the Altes chirplet $u(t)$ at a given value of parameter set $\{\omega_0,\omega_c,\lambda\}$
	for the representation eq.~(\ref{newAltes2}). The parameters are each swept in a dense grid over their domains. For each instance, the delay spread $\tau$ and bandwidth $B$ of the chirplet are charted. It can be seen that certain parameterizations fall on an efficient frontier, representing realizations with maximized time-frequency localization.}
	\label{efficientFig}
\end{figure}

In Section~\ref{chirpRate}, it was seen that investigation of $\lambda\in(0,1)$ is sufficient. All results derived apply then also for $1/\lambda$. Furthermore, as outlined in Section~\ref{examples}, we consider unit sampling and need only consider $\omega_0\in(0,\pi)$ and $\omega_c\in(0,\pi]>\omega_0$. The results of a dense parameter sweep within these domains are presented in Fig.~\ref{efficientFig}. This figure shows the measured delay spread and bandwidth for each grid point, i.e. for multiple realizations of the Altes chirplet at distinct parameter settings.

It can be immediately seen that there is an efficient frontier to the lower left, along which there is an optimal trade-off between time-localization and frequency-localization. Altes chirplets away from this Frontier are inefficient, as they  could be tuned to have smaller bandwidth without increasing the delay-spread and vice-versa. 

We examine the Altes parameters when moving down this curve from upper left to lower right. In all cases we find that $\lambda<1/2$, a loose parameter bound for the efficient frontier. This is because the Altes chirplet in eq.~(\ref{newAltes}) is ill-behaved near $\lambda=1$ and is certainly not efficiently localized there. The effects of this ill-behavior begin to disappear for $\lambda<1/2$. 

At the upper left part of the frontier, $\omega_0>\pi/2$ and $B<\pi/4$, and the waveform becomes frequency localized (narrowband). There is little chirping possible, and in this case, we are dealing with a more conventional wavelet, rather than a chirplet.

Moving down the curve to the right, we see that the cutoff frequency quickly converges to the Nyquist rate, $\omega_c\rightarrow\pi$, while the center frequency slides down to $\omega_o\rightarrow\pi/2$, as the Altes chirplet becomes less frequency localized. Finally, the efficient frontier flattens out at $B=3\pi/4$. This is the point of minimum delay spread, which occurs for chirplets with $\{\omega_0,\omega_c\} = \{\pi/2,\pi\}$. This is consistent with equation (\ref{BW}). Table~\ref{effPar} summarizes the efficient parameterizations discussed in this section, which maximize time-frequency localization.

\subsection{Inefficiently Localized Chirplets}
At this point, we could call it a day, having identified parameterizations of the Altes Chirplet for efficient time-frequency localization. However, it appears that nature values chirping flexibility  above localization-efficiency when it comes to ubiquitous log-periodicities. 

For example, the parameterization in Fig.~\ref{example1} that Altes used to model bat chirps has a high ratio of cutoff to center frequency $\omega_c/\omega_0$ (ie. high bandwidth) and is not frequency localized. However, it is also not sufficiently time-localized to lie on the efficient frontier of Fig.~\ref{efficientFig}. 

More generally, we can say that a) significant chirping occurs away from the frequency localized regime, such that there is a sufficient range of frequencies over which to sweep; and similarly, b) the chirplet must be long enough lived to  to afford numerous oscillations of the log periodicity. Excessive time localization impedes desirable chirping. 

Experience and research has taught us that the chirp waveforms encountered in practice, such as those used by animals for echolocation, as well as those encountered in the signatures of discrete-scale invariance such as fracture dynamics, exhibit \emph{inefficient} time-frequency localization. In simpler words, we seek chirplets with relatively wide bandwidths and long delay spreads, whose parameterizations are selected away from the efficient frontier of Fig.~\ref{efficientFig}. This opens up interesting questions on possible improvements that have remained unearthed until now.
We leave this problem for future analysis.

\begin{table}[t]
\begin{center}
\caption{Parameterizations of the Altes Chirplet for Efficient Time-Frequency Localization}
\label{effPar}
\begin{tabular}{l|c|c|c}
&$\omega_0$&$\omega_c$ or $B$&$\lambda$\\\hline\hline
Efficient Frequency&
$\omega_0>\frac{\pi}{2}$&
$\omega_c<\pi$&$\lambda<\frac{1}{2}$\\
Localization&&
$(B<\frac{\pi}{4})$&
$\Leftrightarrow$\\
 (minimal chirping)&&&$\lambda>2$\\[1ex]\hline
Efficient Time-Freq. &$\omega_0>\frac{\pi}{2}$&$\omega_c=\pi$&$\lambda<\frac{1}{2}$\\
Localization&&$(\frac{\pi}{4}<B<\frac{3\pi}{4})$&$\Leftrightarrow$\\(the frontier)&&&$\lambda>2$\\[1ex]\hline
Efficient Time&$\omega_0=\frac{\pi}{2}$&$\omega_c=\pi$&$\lambda<\frac{1}{2}$\\Localization&&$(B=\frac{3\pi} {4})$&$\Leftrightarrow$\\ (min. delay-spread)&&&$\lambda>2$\\[1ex]\hline
\end{tabular}
\end{center}
\end{table}

\subsection{Discrete-Time Implementation}\label{dti}
In specifying which parameterizations to use, we start by fixing cutoff frequency $\omega_c=\pi$. In Section~\ref{tfloc}, it was found that this is required for efficient localization. Now that inefficiently localized chirps are also in consideration, there remains a good reason to fix $\omega_c=\pi$. In order to minimize the complexity of the discrete-time implementation, the chirplet should be represented by a sampling that is as sparse as possible, such that the wavelet dynamics are fully captured without aliasing. Assuming a unit sampling interval (a sampling frequency of $\omega_s=2\pi$), this translates directly to the requirement that the chirplet cutoff equals the Nyquist sampling rate  $\omega_c=\omega_{\textrm{Nyq}}=\pi$. This, together with Section~\ref{tfloc}, requires that $\omega_0<\frac{\pi}{4}$ (i.e. $B>\frac{3\pi}{4}$) for frequency delocalization i.e. such that wideband chirping can occur.

\begin{figure}[t]
\centering
    \includegraphics[width=7.5cm, totalheight=6cm,angle=0]{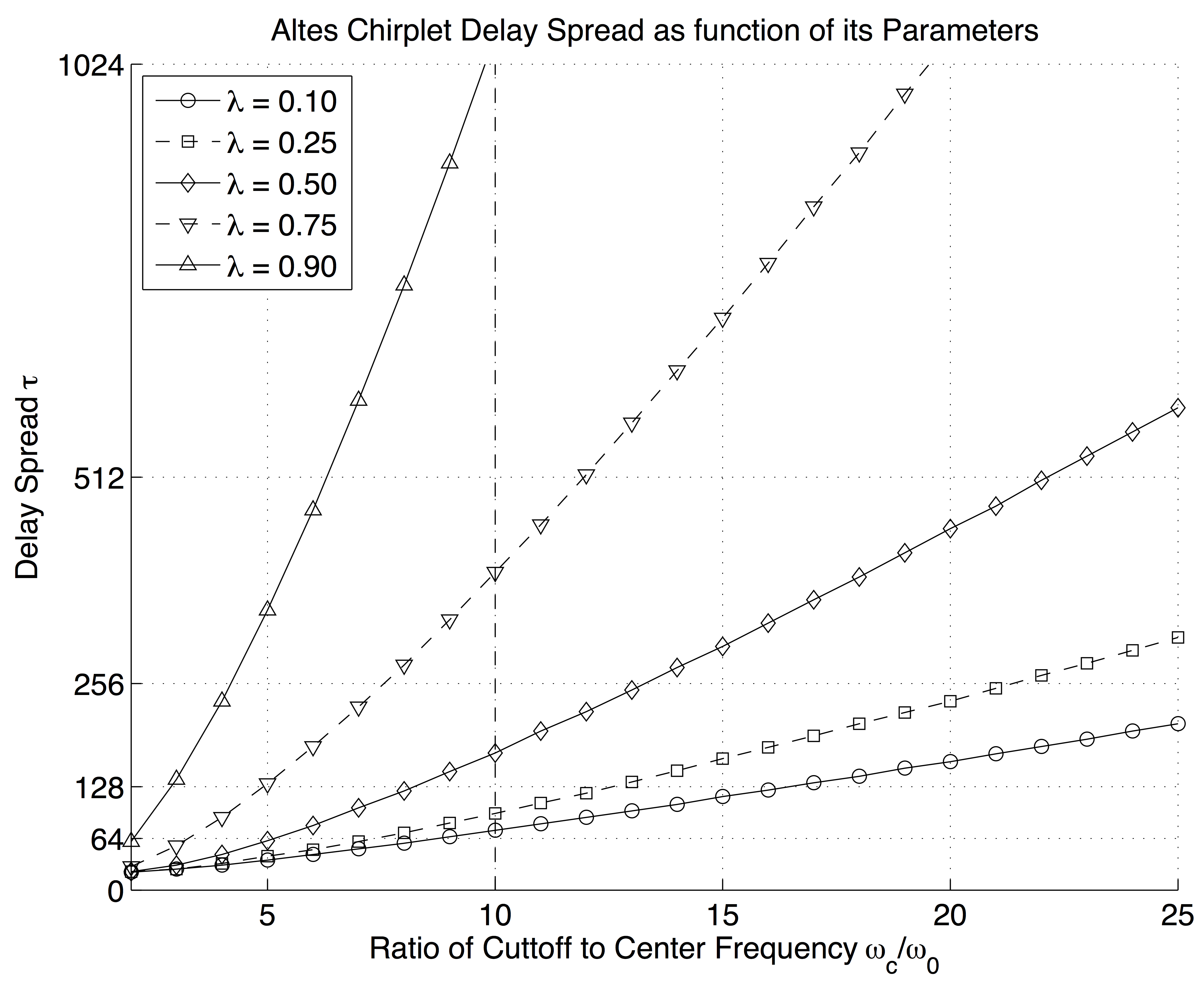}
    	\caption{Altes chirplet delay spread $\tau$ as a function of the parameters $\{\omega_0,\lambda\}$ for inefficiently localized chirps. The cutoff frequency is fixed at $\omega_c=\pi$. At these settings, the delay spread \emph{increases} more or less linearly with bandwidth, which increases complexity of implementation. Note that increasing $\omega_0/\omega_c$ is used as a proxy for increasing bandwidth as per equation (\ref{BW}).
	The vertical dotted line shows by example that a 1024-point Fourier Transform would be the minimum size required to implement all chirplets with $\lambda<0.9$, $\omega_0>\frac{\pi}{10}$.}
    	\label{delaySpread}
\end{figure}

Before moving on to chirp-rate $\lambda$, we take a short aside on complexity. With critical-rate unit-sampling, the number of time-samples required to represent a chirplet is equal to the delay spread $\tau$ since there is approximately no energy outside this window, as described in Section~\ref{dsp}. Now consider Fig.~\ref{delaySpread}, in which it seen that, counterintuitively, the delay-spreads of wideband Altes chirplets \emph{increase} more or less linearly with bandwidth (i.e. when parameterized for inefficient localization). This puts a burden on complexity, which  must be bounded.  To generate the time-domain waveform using the Inverse Fast Fourier Transform (IFFT) in eq.~(\ref{ut}), we would select a transform size that is large enough to capture the full delay spread. Since the IFFT is implemented cost-effectively when its size is an integer power of two, the $y$-axis abscissa of Fig.~\ref{delaySpread} are labeled dyadically, directly allowing the choice of transform size for discrete-time implementation. Given bounds on the Altes parameter set, we can directly select the appropriate IFFT transform size from the chart.

Recall from Section~\ref{scalInv} that the Hyperbolic Chirplet Transform should be computable over a range of chirp-rate values for $0<\lambda<1$ (or equivalently $\lambda>1$ for accelerating rather than decelerating chirps). In addition, as described above, in aiming for wideband, chirping $\omega_c/\omega_0$ should be large. Examining Fig.~\ref{delaySpread}, it appears that, for a fixed delay-spread, these requirements compete for complexity, and a trade-off must be reached. For example, selecting $\frac{\omega_c}{\omega_0}=10$ and the chirp rate  $\lambda=0.75$ would require an FFT size of at least 512 for alias-free computation of the Altes time-domain chirplet $u(t)$ in eq.~(\ref{ut}). This complexity analysis above is crucial for choosing the Fourier transform size in discrete-time implementation of the Altes chirplet.

\begin{figure}[t]
\centering
    \includegraphics[width=7.5cm, totalheight=6cm,angle=0]{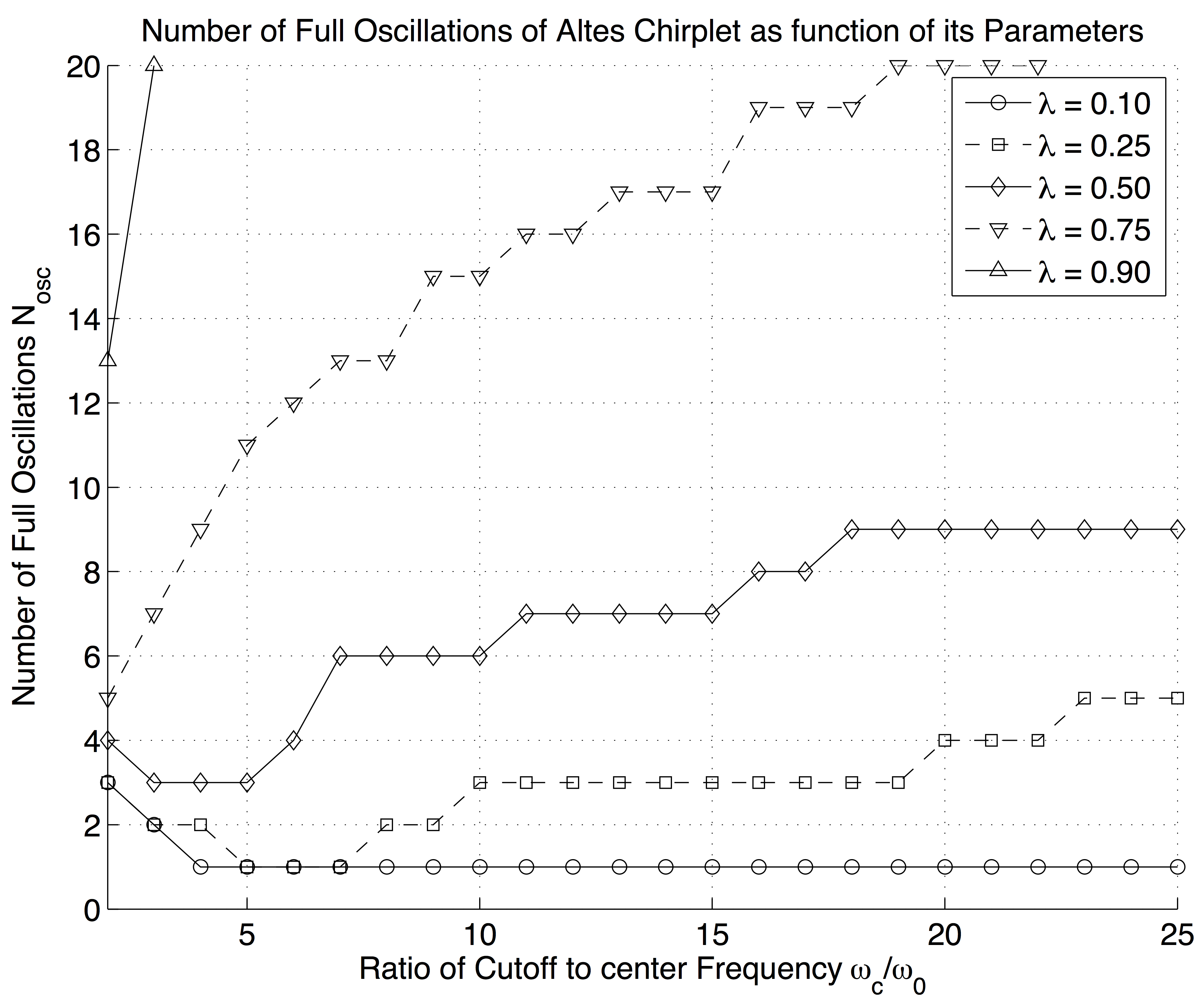}
    	\caption{This chart shows the number of full time-domain oscillations of the Altes chirplet within the limits of its delay spread, as a function of its parameters $\{\omega_0,\omega_c,\lambda\}$. The settings are the same as those used in Fig.~\ref{delaySpread}. For $0 < \lambda <1$, the larger $\lambda$, the larger the number of oscillations, which scales approximately as $-1/\ln(\lambda)$.}
	\label{numOscillation}
\end{figure}

\subsection{Number of Chirplet Oscillations: Discrete Scale Invariance}
We can place bounds on chirping parameter $\lambda$ by examining the number of oscillations of the chirplet. Consider the chirplets in figures~\ref{example1}-\ref{example2}. The first example appears to have about two full oscillations of the waveform, while the second has about six. We can put more precision on this by counting the number of oscillations of the Altes chirplet over a relevant parameter sweep. Repeating the sweep of Fig.~\ref{delaySpread}, the chart of Fig.~\ref{numOscillation} shows the measured number of full oscillations of the chirplet, rather than the measured delay spread. It can be seen that the number of oscillations becomes large for values of chirp-rate close to unity. This is relevant for two main reasons.

Firstly, the well known wavelets such as Meyer, Shannon, Gauss, Morelet, Mexican Hat, etc. all have a limited number of oscillations over their region of support (or within their delay spread, for infinite support wavelets). It has been found that a low number of oscillations, from two or three up to a dozen or so, has given useful results in their application domains, such as data compression and signal analysis. This is because different frequencies and scales can be analyzed using a small number of oscillations, by appropriate dilations. It is purported that this should also hold for the Altes chirplet. 

A more compelling reason stems from the concept of discrete scale invariance introduced in Section~\ref{motivation}. As explained there, signals which are log-periodic are discrete scale invariant, or self-similar. Such signals are invariant under a discrete subset of dilatations. Our Altes chirplet is not perfectly discrete scale invariant, because of its time domain envelope, which forces time-localization, and hence usefulness as a wavelet. However, the chirplet can still be used to detect discrete scale invariance\footnote{The \emph{discrete} scale invariance of a log periodic signal discussed here is different to the full scale invariance of the HCT discussed in Section~\ref{scalInv}.} by its pseudo log-periodicity. The ratio of intervals between the peaks of its oscillation is constant, as in genuine log-periodic signals. This has the subtle advantage that  only a few oscillations of the chirplet are necessary for LP-detection, since further oscillations only detect extensions of the log-periodicity at higher or lower (discrete) scales. More simply, we do not need a large number of chirplet oscillations to detect a log-periodicity. A few cycles will do, plus appropriate dilations and scaling, as in the 
hyperbolic chirplet transform. 

The implications of this become apparent in Fig.~\ref{numOscillation}, which suggests that we need a tighter upper bound than $\lambda<1$. In order to 
remain below 20 oscillations of the chirp, we can retain all of the previous parameter bounds but need to tighten the upper chirp-rate bound to $\lambda<0.75$. 

In addition, we also propose a \emph{lower} bound on the useful number of oscillations. Log-periodicity can occur spuriously in noisy data and it is desirable to avoid falsely reporting such events as significant. 
A deep study on the statistics of random-walk data (integrated noise) has shown that the most likely number of spurious oscillations which occur is 1.5, and that 2.5  can occur with a likelihood as high as 10\% over many realizations \cite{huang2000artifactual}. 
We therefore suggest that usefully parameterized Altes chirplets will have at least 2 oscillations, implying a practical lower bound $\lambda>0.25$ from Fig.~\ref{numOscillation}.

\begin{table}[t]
\begin{center}
\caption{Parameter Bounds on the Altes Chirplet for Discrete-Time Implementation and use in the Hyperbolic Chirplet Transform}
\label{discretePar}
\def\arraystretch{1.3}%
\begin{tabular}{l|c|l}
Parameter&Value(s)&Comment\\\hline\hline
Center Frequency $\omega_0$&$\omega_0<\frac{\pi}{4}$&Wideband Chirping\\[1ex]\hline
Cutoff Frequency $\omega_c$&$\omega_c=\omega_{\textrm{Nyq}}=\pi$&Critical sampling with\\
		&&unit sampling interval\\[1ex]\hline
Chirp-Rate $\lambda$&$1/4<\lambda<3/4$&Bounded Number\\
		&($\Leftrightarrow4/3<\lambda<4$)&of oscillations\\[1ex]\hline
Fourier Transform&Use Fig~\ref{delaySpread}&$N_\textrm{FFT}$=512 will work\\
Size $N_\textrm{FFT}$&to select&for most cases\\[1ex]\hline
\end{tabular}
\end{center}
\end{table}

\begin{figure*}[t]
\centering
\includegraphics[width=15cm,totalheight=11.5cm, angle=0]{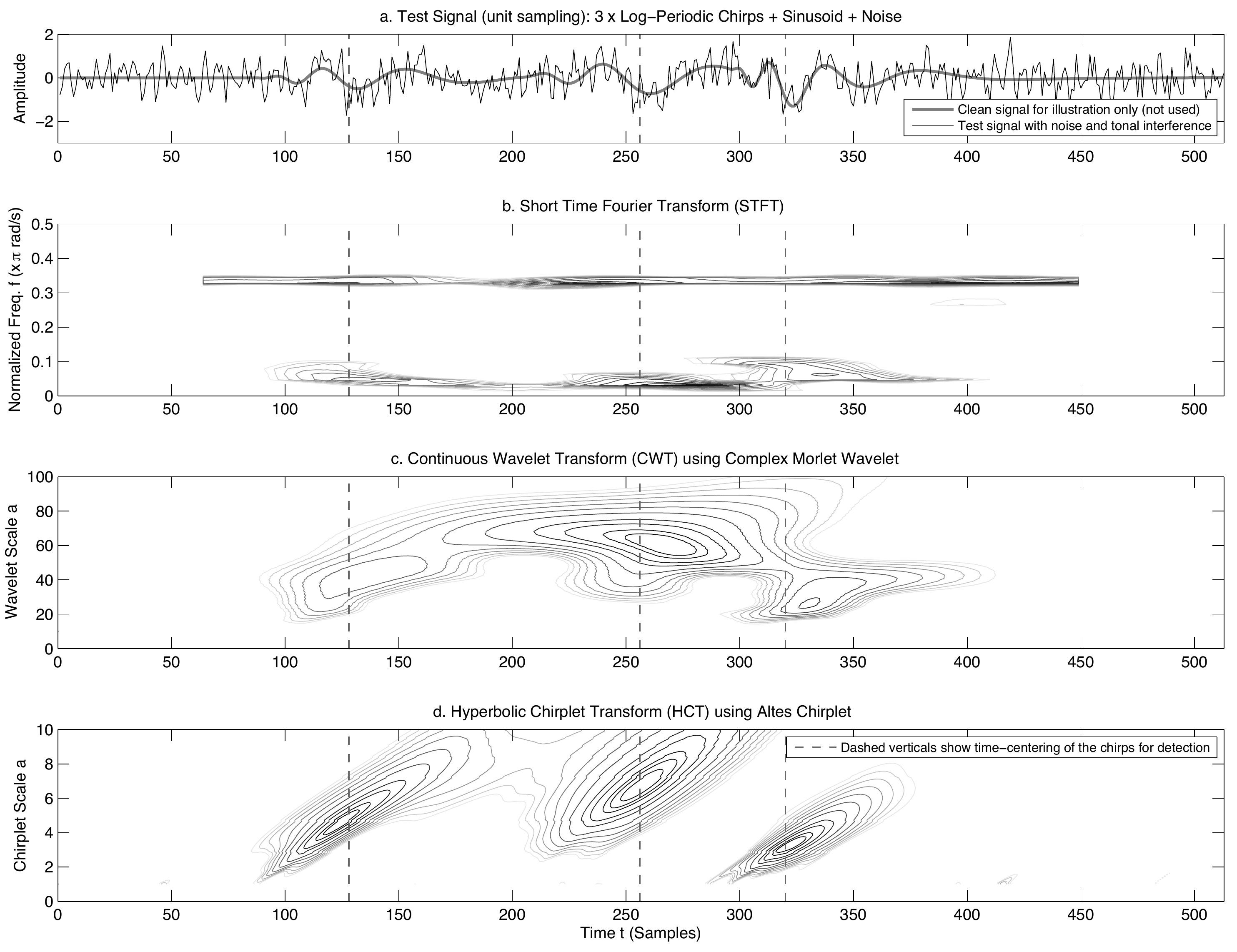}
    	\caption{Comparison of transform analyses for the detection of log-periodicity in a synthetic signal. Subplot a) shows the test signal: three log-periodic chirps and a sinusoid buried in noise. For illustration, the chirps are also shown without noise and tonal interference, and their centers of energy ($\tau_0$ from Fig.~\ref{magfig}b) are marked with dashed verticals. The remaining subplots are log-magnitude contour charts of the transform outputs: b) Short Time Fourier Transform (STFT) spectrogram of the test signal using a 128-point Fourier Transform and Hamming window  c) Continuous Wavelet Transform (CWT) scalogram using the complex Morlet wavelet and d) Hyperbolic Chirplet Transform (HCT) scalogram of the test signal using an Altes chirplet. It can be seen that the HCT isolates the log periodic chirps for detection in the time-scale plane. The formation of linear ridges in the HCT output confirms the presence of log-periodicity.}
    	\label{synthEx}
\end{figure*}

\subsection{Summary of Altes Chirplet Parameter Selection}
In this Section, it has been shown how to parametrize the Altes chirplet for efficient time-frequency localization. However, it has also been noted that localization-efficiency should be sacrificed for chirping flexibility, and instead we seek parameters that allow wide-band chirping for a selection of chirp-rates with minimized sampling rate and Fourier transform size. 
A numerical examination of these properties over the parameter space leads to the conclusions of Table~\ref{discretePar} for parameterizing the Altes chirplet. The snippet of \texttt{\textsc{Matlab$^\textrm{TM}$}} below shows that the chirplet can be implemented in a straight-forward manner in a software environment, given the selected parameters.
\begin{framed}{\footnotesize
\begin{verbatim}
function [U,u] = altesChirplet(w0,wc,lam,Nfft)
%% Altes Chirplet time domain u & freq domain U

Kc = 0.01;                        
Np = (Nfft/2)+1;				
wp = linspace(0,pi,Np);   % positive frequencies	
kc = -log(Kc)/(log(wc/w0)^2);	
Mwp = exp(-kc*(log(wp/w0).^2));	
Qwp = exp(2*pi*j*log(wp)/log(lambda));
U = Mwp.*Qwp;
Upad = [0 U(2:end) zeros(1,(Nfft/2)-1)]; 
u = ifft(Upad);
\end{verbatim}}
\end{framed}

\section{Empirical results using Synthetic Data}\label{empirical}

In order to help demonstrate the applicability of the Altes wavelet to detection of log-periodicities via the Hyperbolic Chirplet Transform, we have constructed a synthetic test signal in which three log-periodic chirps are buried in white noise, along with a sine wave. The resulting signal for analysis is shown in Fig.~\ref{synthEx}a, having a signal to noise ratio (SNR) of 0~dB. For comparison, we have chosen to analyze this signal by three methods, a Short Time Fourier Transform (STFT), a Continuous Wavelet Transform (CWT) using the complex Morlet wavelet, and the Hyperbolic Chirplet Transform using an Altes chirplet. The resulting spectrogram / scalogram outputs are shown in figures~\ref{synthEx}b-\ref{synthEx}d respectively.

In Fig.~\ref{synthEx}b, the first thing to spot is that the STFT picks out a sinusoid, which has been injected at frequency $\frac{\pi}{3}$. This is not detected by the wavelet or chirplet transforms, highlighting the imperative to pick a suitable transform for the problem at hand. Assuming we are only interested in detecting log-periodicities, it is clear that the STFT is relatively poor, as is to be expected from its constant time-resolution at all frequencies. We have implemented an overlap-add STFT using a 128-point Fourier transform with a Hamming window.

Fig.~\ref{synthEx}c shows some improvement by the CWT in detecting the LP-chirps. In Section~\ref{introchirp}, we noted that Morlet wavelets have been reported to exhibit superior performance over other wavelets for chirp detection \cite{sejdic2008quantitative}, so we use a Morlet here as the mother wavelet for the CWT. For fairness of comparison with the (complex) Altes chirplet, we use the complex Morlet, which marginally improves localization in the scalogram. Furthermore, since the complex Morlet has much narrower time-support than the Altes chirplet at unit scale, we use a higher set of scales in the CWT than the HCT, such that the time resolution of their scalograms are comparable. It is found that  we need to choose a set of scales approximately $10\times$ those of the HCT, as can be seen from the $y$-axes of figures~\ref{synthEx}c-d. The CWT scalogram seems to suggest, correctly in fact, that there are three bursts of signal activity over the interval. However, while the result is partially clear around the timing and scaling of signal activity (the localisation of the three LP-chirps is quite poor)  it would be a considerable stretch to conclude that this represents a set of log-periodicities. .

Fig.~\ref{synthEx}d is more promising. There is a significant increase in sharpness identifying the timing and scaling of bursty signals buried in the noise. But of greater interest are the three linear ridges, which sweep upwards from left to right. The HCT has captured the benefits of linearizing chirps  in the time-scale plane, as identified in \cite{mannspieconference}\cite{yin2002fast}, and has correctly isolated the placement of the chirps as synthesized in the artificial test-signal. This is despite the high levels of noise and tonal-interference superimposed on the chirps to be detected. 

In parameterizing the Altes chirplet in this example, we have kept within the bounds of Table~\ref{discretePar}, selecting $\{\omega_0,\omega_c,\lambda\} = \{\frac{\pi}{5},\pi,\frac{1}{2}\}$. In fact, many systems exhibiting discrete scale invariance have a preferred scaling ratio $\lambda=2$ ($\Leftrightarrow \lambda=\frac{1}{2}$). For example in \cite{sornette1998discrete}, it has been observed that the mean field value of $\lambda=2$ is obtained by taking an Ising or Potts model on a hierarchical lattice in the limit of an infinite number of neighbors. Also, we have seen that chirplet parameter $\lambda=\frac{1}{2}$ lies directly between the limits of applicability we found for the HCT $\frac{1}{4}<\lambda<\frac{3}{4}$. We surmise that $\lambda=\frac{1}{2}  \Leftrightarrow \lambda=2$ is a natural choice for the chirp-rate. We find that, when making this selection for our mother wavelet, the HCT will nevertheless succeed in detecting chirps generated with different values of $\lambda$. This is because $\lambda$ is a scale ratio for the distance between successive peaks in a log-periodic waveform, while $a$ is the scale ratio within the HCT that serves as a dilation factor, stretching the analyzing waveform to find log-periodicities at other chirp rates. This is all to say that fixing $\lambda =  \frac{1}{2}$ in the HCT is not seen to limit its use for more general LP-detection. 

It is worth noting that the basic analyzing waveforms for the STFT (a windowed sinusoid) and the CWT (the complex Morlet) are both symmetric in time about their center. However, the Altes Chirplet is skewed (Fig.~\ref{magfig}b), which moves the center of energy when analyzing at different scales. For consistency with the CWT, the Altes chirplet is centered on its energy peak (ie. by setting $\tau_0\stackrel{!}{=}0$ in Fig.~\ref{magfig}b) for HCT implementation.  

To conclude, we see that there is promise in the use of the Altes chirplet and the HCT for improving our ability to detect log-periodic signatures in noisy signals. Looking forward, our research is taking us down a more applied route than the theoretical framework of the current study, and there are certainly  many real-world applications where the value of the methodology can be quantified more precisely. Such application is beyond the scope of this introductory paper.

\section{Conclusion}\label{conclusion}
Building on the excellent sonar waveform designs of R.A. Altes from the 1970s, this article has taken the step to make his work both more accessible to the signal processing community, and more widely applicable in the context of wavelet transform analysis. To achieve the former, a reparameterization allows simple specification of a family of chirplets in terms of bandwidth, center frequency, and chirp-rate. It is demonstrated that these wavelets are admissible, infinitely regular, have infinite vanishing moments, and furthermore, deliver scale invariance when implemented in a continuous wavelet transform. 

For the latter, it has been shown how to design a useful  parameterization of these chirplets for application in a discrete-time Hyperbiolic Chirplet Transform (HCT). We demonstrate that the HCT facilitates detection of log-periodicity (LP) in a noisy signal by linearizing its scalogram signature, a feat not achievable with other time-frequency techniques such as the short time Fourier transform or the continuous wavelet transform. These theoretical underpinnings for LP-detection can form the basis for applied research in multidisciplinary settings, particularly where there is an imperative to diagnose criticality, and forecast rupture/failure in complex systems.

\bibliographystyle{IEEEtran}
\bibliography{IEEEabrv,altesWavelet_Oct2018}

\begin{thebibliography}{10}
\providecommand{\url}[1]{#1}
\csname url@samestyle\endcsname
\providecommand{\newblock}{\relax}
\providecommand{\bibinfo}[2]{#2}
\providecommand{\BIBentrySTDinterwordspacing}{\spaceskip=0pt\relax}
\providecommand{\BIBentryALTinterwordstretchfactor}{4}
\providecommand{\BIBentryALTinterwordspacing}{\spaceskip=\fontdimen2\font plus
\BIBentryALTinterwordstretchfactor\fontdimen3\font minus
  \fontdimen4\font\relax}
\providecommand{\BIBforeignlanguage}[2]{{%
\expandafter\ifx\csname l@#1\endcsname\relax
\typeout{** WARNING: IEEEtran.bst: No hyphenation pattern has been}%
\typeout{** loaded for the language `#1'. Using the pattern for}%
\typeout{** the default language instead.}%
\else
\language=\csname l@#1\endcsname
\fi
#2}}
\providecommand{\BIBdecl}{\relax}
\BIBdecl

\bibitem{altes1970n1}
R.~A. Altes and E.~L. Titlebaum, ``Bat signals as optimally {D}oppler tolerant
  waveforms,'' \emph{The Journal of the Acoustical Society of America},
  vol.~48, no.~4, pp. 1014--1020, 1970.

\bibitem{altes1973n2}
R.~Altes, ``Some invariance properties of the wide-band ambiguity function,''
  \emph{The Journal of the Acoustical Society of America}, vol.~53, p. 1154,
  1973.

\bibitem{altes1975n2}
R.~A. Altes, ``Sonar for generalized target description and its similarity to
  animal echolocation systems,'' \emph{The Journal of the Acoustical Society of
  America}, vol.~59, no.~1, pp. 97--105, 1975.

\bibitem{altes1975n4}
R.~A. Altes and W.~D. Reese, ``Doppler-tolerant classification of distributed
  targets -- a bionic sonar,'' \emph{{IEEE} Trans. Aerosp. Electron. Syst.},
  vol.~11, no.~5, pp. 708--724, 1975.

\bibitem{altes1977n2}
D.~P. Skinner, R.~A. Altes, and J.~D. Jones, ``Broadband target classification
  using a bionic sonar,'' \emph{The Journal of the Acoustical Society of
  America}, vol.~62, no.~5, pp. 1239--1246, 1977.

\bibitem{flandrin1990generalized}
P.~Flandrin, F.~Magand, and M.~Zakharia, ``Generalized target description and
  wavelet decomposition [sonar],'' \emph{Acoustics, Speech and Signal
  Processing, IEEE Transactions on}, vol.~38, no.~2, pp. 350--352, 1990.

\bibitem{sornette1998discrete}
D.~Sornette, ``Discrete-scale invariance and complex dimensions,''
  \emph{Physics reports}, vol. 297, no.~5, pp. 239--270, 1998, extended version
  available online: http://xxx.lanl.gov/abs/cond-mat/9707012.

\bibitem{Daubechies92}
I.~Daubechies, \emph{Ten lectures on wavelets}.\hskip 1em plus 0.5em minus
  0.4em\relax SIAM, 1992, vol.~61.

\bibitem{Mallat98}
S.~Mallat, \emph{A wavelet tour of signal processing: the sparse way}.\hskip
  1em plus 0.5em minus 0.4em\relax 3rd ed. 2009, Academic Press, 1998.

\bibitem{sornette2009stock}
D.~Sornette, \emph{Why stock markets crash: critical events in complex
  financial systems}.\hskip 1em plus 0.5em minus 0.4em\relax Princeton
  University Press, 2017.

\bibitem{Johan-sornetterupt00}
A.~Johansen and D.~Sornette, ``Critical ruptures,'' \emph{The European Physical
  Journal B-Condensed Matter and Complex Systems}, vol.~18, no.~1, pp.
  163--181, 2000.

\bibitem{mann1991chirplet}
S.~Mann and S.~Haykin, ``The chirplet transform: A generalization of gabor’s
  logon transform,'' in \emph{Vision Interface}, vol.~91, 1991, pp. 205--212.

\bibitem{mann1995chirplet}
------, ``The chirplet transform: Physical considerations,'' \emph{Signal
  Processing, IEEE Transactions on}, vol.~43, no.~11, pp. 2745--2761, 1995.

\bibitem{lu2008fast}
Y.~Lu, E.~Oruklu, and J.~Saniie, ``Fast chirplet transform with fpga-based
  implementation,'' \emph{Signal Processing Letters, IEEE}, vol.~15, pp.
  577--580, 2008.

\bibitem{mannspieconference}
S.~Mann and S.~Haykin, ``{The Adaptive Chirplet: An Adaptive Wavelet Like
  Transform},'' \emph{SPIE, 36th Annual International Symposium on Optical and
  Optoelectronic Applied Science and Engineering}, 21-26 July 1991.

\bibitem{yin2002fast}
Q.~Yin, S.~Qian, and A.~Feng, ``A fast refinement for adaptive gaussian
  chirplet decomposition,'' \emph{Signal Processing, IEEE Transactions on},
  vol.~50, no.~6, pp. 1298--1306, 2002.

\bibitem{sejdic2008quantitative}
E.~Sejdic, I.~Djurovic, and L.~Stankovic, ``Quantitative performance analysis
  of scalogram as instantaneous frequency estimator,'' \emph{Signal Processing,
  IEEE Transactions on}, vol.~56, no.~8, pp. 3837--3845, 2008.

\bibitem{peng2011polynomial}
Z.~Peng, G.~Meng, F.~Chu, Z.~Lang, W.~Zhang, and Y.~Yang, ``Polynomial chirplet
  transform with application to instantaneous frequency estimation,''
  \emph{Instrumentation and Measurement, IEEE Transactions on}, vol.~60, no.~9,
  pp. 3222--3229, 2011.

\bibitem{yang2013multicomponent}
Y.~Yang, W.~Zhang, Z.~Peng, and G.~Meng, ``Multicomponent signal analysis based
  on polynomial chirplet transform,'' \emph{IEEE Transactions on Industrial
  Electronics}, vol.~60, no.~9, pp. 3948--3956, 2013.

\bibitem{le2004hyperbolic}
K.~N. Le, K.~P. Dabke, and G.~K. Egan, ``Hyperbolic wavelet family,''
  \emph{Review of scientific instruments}, vol.~75, no.~11, pp. 4678--4693,
  2004.

\bibitem{abry2012hyperbolic}
P.~Abry, M.~Clausel, S.~Jaffard, S.~Roux, and B.~Vedel, ``Hyperbolic wavelet
  transform: an efficient tool for multifractal analysis of anisotropic
  textures,'' \emph{arXiv preprint arXiv:1210.1944}, 2012.

\bibitem{flandrin2001time}
P.~Flandrin, ``Time frequency and chirps,'' in \emph{Aerospace/Defense Sensing,
  Simulation, and Controls}.\hskip 1em plus 0.5em minus 0.4em\relax
  International Society for Optics and Photonics, 2001, pp. 161--175.

\bibitem{chassande1999time}
E.~Chassande-Mottin and P.~Flandrin, ``On the time--frequency detection of
  chirps,'' \emph{Applied and Computational Harmonic Analysis}, vol.~6, no.~2,
  pp. 252--281, 1999.

\bibitem{poularikas2010transforms}
J.~Bertrand, P.~Bertrand, and J.-P. Ovarlez, \emph{Transforms and applications
  handbook: Chapter 12 --- The Mellin Transform}, ser. The Electrical
  Engineering Handbook Series, A.~D. Poularikas, Ed.\hskip 1em plus 0.5em minus
  0.4em\relax Florida, USA: CRC Press, 1995.

\bibitem{gluzman2002log}
S.~Gluzman and D.~Sornette, ``Log-periodic route to fractal functions,''
  \emph{Physical Review E}, vol.~65, no. 036142, 2002.

\bibitem{yiou2000dataAdaptive}
P.~Yiou, D.~Sornette, and M.~Ghil, ``Data-adaptive wavelets and multi-scale
  singular-spectrum analysis,'' \emph{Physica D: Nonlinear Phenomena}, vol.
  142, no.~3, pp. 254--290, 2000.

\bibitem{saleur1996discrete}
H.~Saleur, C.~Sammis, and D.~Sornette, ``Discrete scale invariance, complex
  fractal dimensions, and log-periodic fluctuations in seismicity,''
  \emph{Journal of Geophysical Research: Solid Earth (1978--2012)}, vol. 101,
  no.~B8, pp. 17\,661--17\,677, 1996.

\bibitem{wornell1992wavelet}
G.~W. Wornell and A.~V. Oppenheim, ``Wavelet-based representations for a class
  of self-similar signals with application to fractal modulation,''
  \emph{Information Theory, IEEE Transactions on}, vol.~38, no.~2, pp.
  785--800, 1992.

\bibitem{zababakhin1966shock}
E.~Zababakhin, ``Shock waves in layered systems,'' \emph{Zh. Eksp. Teor. Fiz},
  vol.~49, 1966.

\bibitem{novikov1966effects}
E.~Novikov, ``The effects of intermittency on statistical characteristics of
  turbulence and scale similarity of breakdown coefficients,''
  \emph{Dokl.Akad.Nauk SSSR}, vol. 168, no.~6, p. 1279, 1966.

\bibitem{barenblatt1971intermediate}
G.~I. Barenblatt and Y.~B. Zel'dovich, ``Intermediate asymptotics in
  mathematical physics,'' \emph{Russian Mathematical Surveys}, vol.~26, no.~2,
  pp. 45--61, 1971.

\bibitem{jona1975renormalization}
G.~Jona-Lasinio, ``The renormalization group: A probabilistic view,'' \emph{Il
  Nuovo Cimento B Series 11}, vol.~26, no.~1, pp. 99--119, 1975.

\bibitem{nauenberg1975scaling}
M.~Nauenberg, ``Scaling representation for critical phenomena,'' \emph{Journal
  of Physics A: Mathematical and General}, vol.~8, no.~6, p. 925, 1975.

\bibitem{van1976phase}
T.~Niemeijer and J.~M.~J. Van~Leeuwen, \emph{Phase Transitions and Critical
  Phenomena}, C.~Domb and M.~S. Green, Eds.\hskip 1em plus 0.5em minus
  0.4em\relax London Academic Press, 1976.

\bibitem{kapitulnik1983self}
A.~Kapitulnik, A.~Aharony, G.~Deutscher, and D.~Stauffer, ``Self similarity and
  correlations in percolation,'' \emph{Journal of Physics A: Mathematical and
  General}, vol.~16, no.~8, p. L269, 1983.

\bibitem{doucot1986first}
B.~Doucot, W.~Wang, J.~Chaussy, B.~Pannetier, R.~Rammal, A.~Vareille, and
  D.~Henry, ``First observation of the universal periodic corrections to
  scaling: Magnetoresistance of normal-metal self-similar networks,''
  \emph{Physical review letters}, vol.~57, no.~10, p. 1235, 1986.

\bibitem{bessis1987mellin}
D.~Bessis, J.~Fournier, G.~Servizi, G.~Turchetti, and S.~Vaienti, ``Mellin
  transforms of correlation integrals and generalized dimension of strange
  sets,'' \emph{Physical Review A}, vol.~36, no.~2, p. 920, 1987.

\bibitem{fournier1989singularity}
J.-D. Fournier, G.~Turchetti, and S.~Vaienti, ``Singularity spectrum of
  generalized energy integrals,'' \emph{Physics Letters A}, vol. 140, no.~6,
  pp. 331--335, 1989.

\bibitem{sornette1995complex}
D.~Sornette and C.~G. Sammis, ``Complex critical exponents from renormalization
  group theory of earthquakes: Implications for earthquake predictions,''
  \emph{Journal de Physique I}, vol.~5, no.~5, pp. 607--619, 1995.

\bibitem{saleur1996renormalization}
H.~Saleur, C.~G. Sammis, and D.~Sornette, ``Renormalization group theory of
  earthquakes,'' \emph{Nonlinear Processes in Geophysics}, vol.~3, no.~2, pp.
  102--109, 1996.

\bibitem{anifrani1995universal}
J.-C. Anifrani, C.~Le~Floc'h, D.~Sornette, and B.~Souillard, ``Universal
  log-periodic correction to renormalization group scaling for rupture stress
  prediction from acoustic emissions,'' \emph{Journal de Physique I}, vol.~5,
  no.~6, pp. 631--638, 1995.

\bibitem{sornette1996stock}
D.~Sornette, A.~Johansen, and J.-P. Bouchaud, ``Stock market crashes,
  precursors and replicas,'' \emph{Journal de Physique I}, vol.~6, no.~1, pp.
  167--175, 1996.

\bibitem{sornette2001significance}
D.~Sornette, A.~Johansen \emph{et~al.}, ``Significance of log-periodic
  precursors to financial crashes,'' \emph{Quantitative Finance}, vol.~1,
  no.~4, pp. 452--471, 2001.

\bibitem{bessis1983complex}
D.~Bessis, J.~Geronimo, and P.~Moussa, ``Complex spectral dimensionality on
  fractal structures,'' \emph{Journal de Physique Lettres}, vol.~44, no.~24,
  pp. 977--982, 1983.

\bibitem{derrida1983fractal}
B.~Derrida, L.~De~Seze, and C.~Itzykson, ``Fractal structure of zeros in
  hierarchical models,'' \emph{Journal of Statistical Physics}, vol.~33, no.~3,
  pp. 559--569, 1983.

\bibitem{meurice1995evidence}
Y.~Meurice, G.~Ordaz, and V.~Rodgers, ``Evidence for complex subleading
  exponents from the high-temperature expansion of {D}yson's hierarchical
  {I}sing model,'' \emph{Physical review letters}, vol.~75, no.~25, p. 4555,
  1995.

\bibitem{kutnjak1996sandpile}
B.~Kutnjak-Urbanc, S.~Zapperi, S.~Milo{\v{s}}evi{\'c}, and H.~E. Stanley,
  ``Sandpile model on the sierpinski gasket fractal,'' \emph{Physical Review
  E}, vol.~54, no.~1, p. 272, 1996.

\bibitem{vetterli1992wavelets}
M.~Vetterli and C.~Herley, ``Wavelets and filter banks: Theory and design,''
  \emph{Signal Processing, IEEE Transactions on}, vol.~40, no.~9, pp.
  2207--2232, 1992.

\bibitem{altes1980models}
R.~A. Altes, ``Models for echolocation,'' in \emph{Animal sonar systems}.\hskip
  1em plus 0.5em minus 0.4em\relax Springer, 1980, pp. 625--671.

\bibitem{kroszczynski1969pulse}
J.~J. Kroszczynski, ``Pulse compression by means of linear-period modulation,''
  \emph{Proceedings of the IEEE}, vol.~57, no.~7, pp. 1260--1266, 1969.

\bibitem{geraskin2013everything}
P.~Geraskin and D.~Fantazzini, ``Everything you always wanted to know about
  log-periodic power laws for bubble modeling but were afraid to ask,''
  \emph{The European Journal of Finance}, vol.~19, no.~5, pp. 366--391, 2013.

\bibitem{aguiar2014continuous}
L.~Aguiar-Conraria and M.~J. Soares, ``The continuous wavelet transform: moving
  beyond uni-and bivariate analysis,'' \emph{Journal of Economic Surveys},
  vol.~28, no.~2, pp. 344--375, 2014.

\bibitem{huang2000artifactual}
Y.~Huang, A.~Johansen, M.~Lee, H.~Saleur, and D.~Sornette, ``Artifactual
  log-periodicity in finite size data- relevance for earthquake aftershocks,''
  \emph{Journal of Geophysical Research}, vol. 105, p.~25, 2000.

\end{thebibliography}

\end{document}